\documentclass[%
 aip,
 amsmath,amssymb,
 reprint,%
]{revtex4-1}

\usepackage{graphicx}
\usepackage{dcolumn}
\usepackage{bm}
\usepackage{booktabs}
\usepackage[utf8]{inputenc}
\usepackage[T1]{fontenc}
\usepackage{mathptmx}
\usepackage{etoolbox}

\makeatletter
\def\@email#1#2{%
 \endgroup
 \patchcmd{\titleblock@produce}
  {\frontmatter@RRAPformat}
  {\frontmatter@RRAPformat{\produce@RRAP{*#1\href{mailto:#2}{#2}}}\frontmatter@RRAPformat}
  {}{}
}%
\makeatother
\begin{document}

\preprint{AIP/123-QED}

\title[Quantum-assisted tracer dispersion in turbulent shear flow]{Quantum-assisted Lagrangian tracer dispersion in turbulent shear flow}

\author{Julia Ingelmann}
\affiliation{Institute of Thermodynamics and Fluid Mechanics, Technische Universit\"at 
Ilmenau, P.O.Box 100565, D-98684 Ilmenau, Germany}
 
\author{Fabian Schindler}
\affiliation{Institute of Thermodynamics and Fluid Mechanics, Technische Universit\"at 
Ilmenau, P.O.Box 100565, D-98684 Ilmenau, Germany}

\author{Jörg Schumacher}
\homepage{https://www.tu-ilmenau.de/tsm}
\email{joerg.schumacher@tu-ilmenau.de}
\affiliation{Institute of Thermodynamics and Fluid Mechanics, Technische Universit\"at 
Ilmenau, P.O.Box 100565, D-98684 Ilmenau, Germany}

\date{\today}

\begin{abstract}
We present a quantum-assisted generative algorithm for synthetic tracks of Lagrangian tracer particles in a turbulent shear flow. The parallelism and sampling properties of quantum algorithms are used to build and optimize a parametric quantum circuit, which generates a quantum state that corresponds to the joint probability density function of the classical turbulent velocity components, $p(u_1^{\prime}, u_2^{\prime}, u_3^{\prime})$. Velocity samples are drawn by one-shot measurements on the quantum circuit. The hybrid quantum-classical algorithm is validated with two classical methods, a standard stochastic Lagrangian model and a classical sampling scheme in the form of a Markov-chain Monte Carlo approach. We consider a homogeneous turbulent shear flow with a constant shear rate $S$ as a proof of concept for which the velocity fluctuations are Gaussian. The generation of the joint probability density function is also tested on a real quantum device, the 20-qubit IQM Resonance quantum computing platform for cases of up to 10 qubits. Our study paves the way to applications of Lagrangian small-scale parameterizations of turbulent transport in complex turbulent flows by quantum computers. 
\end{abstract}

\maketitle

\section{\label{sec:level1}Introduction}
Quantum computing algorithms demonstrated new ways to classify, approximate and generate functional dependencies in high-dimensional data records in the past decade.\cite{Preskill2018,Deutsch2020} This research found its way into applications in condensed matter physics and quantum chemistry, but also into classical nonlinear physics, such as fluid dynamics. Besides the solution of equations of fluid motion in one- and two-dimensional setups\cite{Bharadwaj2020,Lubasch2020,Succi2023a,Ingelmann2024, Tennie2025, Pfeffer2025}, quantum machine learning algorithms\cite{Biamonte2017} were applied to fluid mechanics, such as quantum reservoir computing \cite{Fujii2017,Chen2020} to build surrogate models of fluid flows.\cite{Pfeffer2023,Ahmed2024} In the following, we want to discuss a simple generative quantum machine learning algorithm for an application to a classical turbulent shear flow. 

The transport of small particles, called tracers, in turbulent flows requires a Lagrangian description.\cite{Yeung2002,Toschi2009} This perspective on (turbulent) flows uses a frame of reference that is attached to each of these individual infinitesimal fluid tracers. The individual tracer particle tracks ${\bm x}(t ;{\bm x}_0)$ follow by 
\begin{align}
    \frac{d{\bm x}}{dt}={\bm u}({\bm x},t)\,,\quad {\bm x}_0 = {\bm x}(t_0)\,
    \label{eq:Lag1}
\end{align}
with the tracer coordinates ${\bm x}=(x_1, x_2, x_3)$ and the (turbulent) velocity vector field ${\bm u}$. Applications of this Lagrangian framework can be found in industrial and environmental contexts; the focus is then often on a quantification of the turbulent dispersion of a cloud of tracer particles, a measure of the mixing properties of the underlying fluid turbulence. Several important applications can be given. For environmental flows, they reach from the long-range transport of microplastics in the ocean \cite{Worm2017,Zhao2025} or volcanic ash spreading in the atmosphere \cite{Ram1991,Woods2010,Arnalds2013} to the dispersion of pollen in wind fields.\cite{Hamaoui2015} The local stirring properties of the flow are probed by the velocity gradient tensor ${\bm\nabla}{\bm u}$ along the Lagrangian tracer tracks ${\bm x}(t)$, assuming that velocity fields are available, which are resolved down to the smallest vortices in the flow.\cite{Meneveau2011,Goetzfried2019} However, in most applications this particle transport has to be modeled, since velocity fields are not available on all scales. The corresponding stochastic models are termed Lagrangian particle dispersion models (LPDM)\cite{Lin2003,Szintai2010,Sun2018}; the advection of tracers by the unresolved fields is described as a stochastic process.\cite{Pope2002} Lagrangian tracer tracks and related statistical moments can also be obtained synthetically by generative machine learning algorithms. This was shown very recently for homogeneous isotropic turbulence by a generative diffusion model.\cite{Li2024} Generative machine learning has been combined in the past years with quantum algorithms to quantum generative machine learning.\cite{Dallaire2018,Ghao2022,Rudolph2022,Kyriienko2024} The corresponding architectures are typically hybrid in nature, which means that classical and quantum algorithm blocks must be combined. These algorithms generated, for example, new images\cite{Huang2021} or random distributions.\cite{Zoufal2019}

In the present work, we want to model the Lagrangian particle dispersion by a quantum-assisted generative algorithm, which uses a direct variational generator (DVG) as a sampler. To this end, a hybrid quantum-classical algorithm is set up\cite{Romero2019,Rudolph2022}, which can generate Lagrangian tracer motion in a synthetic turbulent velocity field with the correct statistical cross correlations between the velocity components and thus model the classical tracer particle dispersion in a turbulent shear flow. Our algorithm is built on a parametric quantum circuit that is optimized in a variational algorithm to build a high-dimensional multi-qubit quantum state. The latter will correspond to a joint probability density function (PDF) of the turbulent velocity components. We will benefit from the quantum parallelism and entanglement, two unique capabilities of quantum algorithms, see also Appendix~\ref{sec:introQC}. 

In our proof-of-concept study synthetic Lagrangian tracer particle tracks are generated by drawing samples, which are nothing else but single-shot runs of the quantum DVG, from this (high-dimensional) distribution. The generated tracer tracks allow us to calculate the mean particle pair dispersion and to determine the different scaling regimes of this essential measure of turbulent mixing with respect to time $t$. The algorithm is tested in a so-called turbulent homogeneous shear flow (HSF), which is characterized by a constant shear rate $S$ and is considered as a first nontrivial extension of the highly idealized homogeneous isotropic turbulence case. This turbulent flow is however still simple enough to have accurate and controllable stochastic results on the particle dispersion for a validation of our quantum computing model. In a HSF, the velocity field is decomposed into a mean flow part, which is a linear flow here, and a turbulent part that varies about the mean flow part, giving rise to
\begin{align}
    {\bm u}({\bm x},t) = U_1(x_3) {\bm e}_1 + {\bm u}^{\prime}({\bm x},t)\quad \mbox{with}\quad U_1(x_3)=Sx_3\,.
    \label{eq:Lag2}
\end{align}
The mean flow $U_1(x_3)$ is time-independent and uni-directional, pointing along the {\em streamwise} $x_1$ direction only. Here, $x_2$ is the {\em spanwise} and $x_3$ the {\em shear} direction. The small-scale turbulent velocity part is denoted ${\bm u}^{\prime}$; this fraction must be modeled by the quantum algorithm. We compare the performance of our algorithm with a stochastic transport model that solves a coupled system of Langevin equations and with a classical Markov-chain Monte Carlo sampling method. The generation of the joint PDF is also implemented on a noisy intermediate scale quantum (NISQ) device, the 20-qubit IQM Garnet Quantum Computer\cite{iqm2024} and compared with ideal statevector simulations that apply the quantum simulation software Qiskit.\cite{qiskit2024}  

The outline of the manuscript is as follows. Section II introduces the DVG algorithm together with some basics of the variational quantum algorithm. Section III reviews the homogeneous shear flow compactly and reports the reconstruction of the joint distributions by the quantum circuit. This is followed by an analysis of the results of the DVG and their comparison with two classical methods. Finally, we give a summary and an outlook in Sec. IV. The appendices summarize background information on quantum computing, stochastic modeling, important scales and parameters of the considered HSF, results for a one-dimensional random walk with IQM Garnet, and the Markov-chain Monte Carlo sampling method for comparison.

\section{Hybrid Direct Variational Generator}
The DVG belongs to the class of Variational Quantum Generators \cite{Romero2019}, the specific notion was termed to the best of our knowledge in ref. \cite{Rudolph2020}. We use a DVG to approximate an $n$-qubit target quantum state optimally. This target quantum state is given by
\begin{equation}
|\Psi^{\ast}\rangle =\sum_{i=0}^{2^{n}-1} a_i |i\rangle  \quad\mbox{with}\quad \|\Psi^{\ast}\|^2_{\cal H}=\sum_{i=0}^{2^{n}-1} |a_i|^2=1\,,
\end{equation}
where $a_i\in \mathbb{C}$ and $|i\rangle$ denotes the $i$-th basis vector of the Hilbert space ${\cal H}=[\mathbb{C}^2]^{\otimes n}$ in bra-ket notation.\cite{Nielsen2010} See again Appendix~\ref{sec:introQC}. Each subspace $\mathbb{C}^2$ is spanned by the two basis vectors $|0\rangle$ and $|1\rangle$. The $2^n$-dimensional complex vector $|\Psi^{\ast}\rangle$ represents the probability amplitudes of a quantum state and is utilized to reconstruct the joint probability density function (PDF) of the turbulent velocity components, \( p(u_1^{\prime}, u_2^{\prime}, u_3^{\prime}) \). More precisely, we focus on the joint PDF \( p(\xi_1, \xi_2, \xi_3) \), where \( \xi_i \) denote the components of vector-valued stochastic noise with the right cross correlations, modeled by the DVG. In the HSF case, we can simplify the joint probability distribution as follows, 
\begin{equation}
p(\xi_1,\xi_2,\xi_3) = p(\xi_1,\xi_3) p(\xi_2)\,,
\label{eq:jpdf}
\end{equation}
since the statistics in the spanwise direction $x_2$ is uncorrelated with the other two space directions which we will further detail in Sec. III. The vector of probability amplitudes ${\bm P}=(|a_0|^2, \dots, |a_{2^n-1}|^2)^T$ is obtained by a repeated measurement of identically prepared quantum systems, the Positive Operator-Valued Measure (POVM). These probability amplitudes have to be mapped to $p(\xi_1, \xi_2, \xi_3)$. For this, a parameterized quantum circuit is trained to generate the wanted target state and hence, can be used as DVG. Subsequently, a one-shot measurement is used to draw a sample from the joint PDF. The algorithm thus uses the unique property of the measurement in a quantum system for the sampling, namely being a random projection on one of the $2^n$ eigenstates of an observable.\cite{Nielsen2010} The standard observable to be used for each of the $n$ qubits is the $z$-component of the spin, the Pauli matrix $Z$. 

\begin{figure}
\centering
\includegraphics[width=0.45\textwidth]{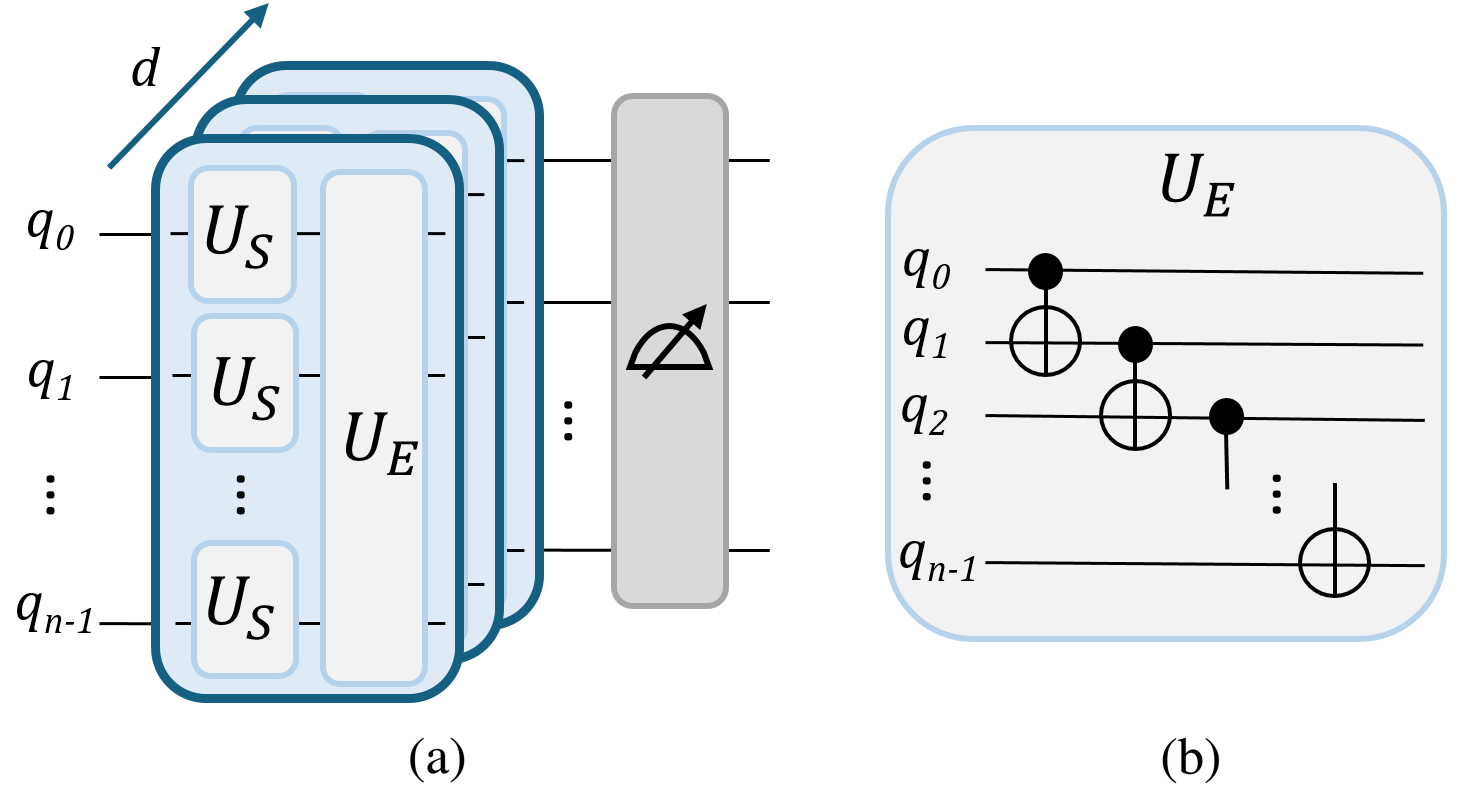}   
\caption{A sketch of (a) the Direct Variational Generator where $U_S$ are single-qubit unitary gates and $U_E$ entanglement gates that connect the $n$ qubits. (b) Detailed unitary representation of entanglement gate $U_E$. In this sketch, it is realized by a sequence of pairwise controlled NOT (CNOT) gates, which successively connect two qubits and thus eventually all $n$ qubits of the $n$-qubit quantum register.}
\label{fig:DVG_Sketch}
\end{figure}

\subsection{Quantum part -- Generation of entangled quantum state }
The DVG is a parametric quantum circuit, composed of single- and two-qubit gates, which performs a unitary transformation of an $n$-qubit quantum register, $|0\rangle^{\otimes n}$, such that a target state $|\Psi^{\ast}\rangle$ is obtained and subsequently measured. We distinguish between single-qubit unitary transformations $U_S$, which comprise qubit rotations and are performed by single-qubit gates, and two-qubit unitary transformations $U_E$, which generate entanglement between qubits and are realized by controlled NOT (CNOT) gates.\cite{Rudolph2020} The single-qubit gates are parameterized by a rotation angle $\theta$, such that $U_S(\theta)$. The $n$-qubit quantum state, which is generated by this multilayer quantum circuit, can be expressed as
\begin{align}
    \vert \Psi \rangle_{\rm gen} = \prod_{j=1}^d U_E^{(j)} \left[\bigotimes_{i=1}^n U_{S,i}^{(j)}\left(\theta_i^{(j)}\right)\right]\, \vert 0 \rangle ^{\otimes n}\,,
\end{align}
where $d$ is the circuit depth parameter and index $j$ runs over the $d$ stacked layers of (identical) quantum gate composition. The circuit depth $d$ determines how good the model can approximate the target state $|\Psi^{\ast}\rangle$. This aspect is connected to the expressibility measure of the circuit.\cite{Holmes2022} The depth is typically limited due to propagating errors caused by decoherent gates on NISQ devices. Figure~\ref{fig:DVG_Sketch}(a) visualizes the general structure of the DVG, see also ref. \cite{Rudolph2020}. In this work, $U_S$ will be rotational $R_Y(\theta)$ gates. This gate performs a rotation of the qubit about the $y$-axis by a rotational angle $\theta/2$. All angles of the parametric quantum circuit with its $d$ layers will be summarized in a parameter vector ${\bm \Theta}$. The unitary $U_E$ is detailed in Fig.~\ref{fig:DVG_Sketch}(b). Due to the connection of qubit $q_0$ with $q_{n-1}$, the present entanglement structure is called linear. \cite{Nakhl2024} Further details will follow below in Sec. III B. 

\begin{figure}
\centering
\includegraphics[width=0.45\textwidth]{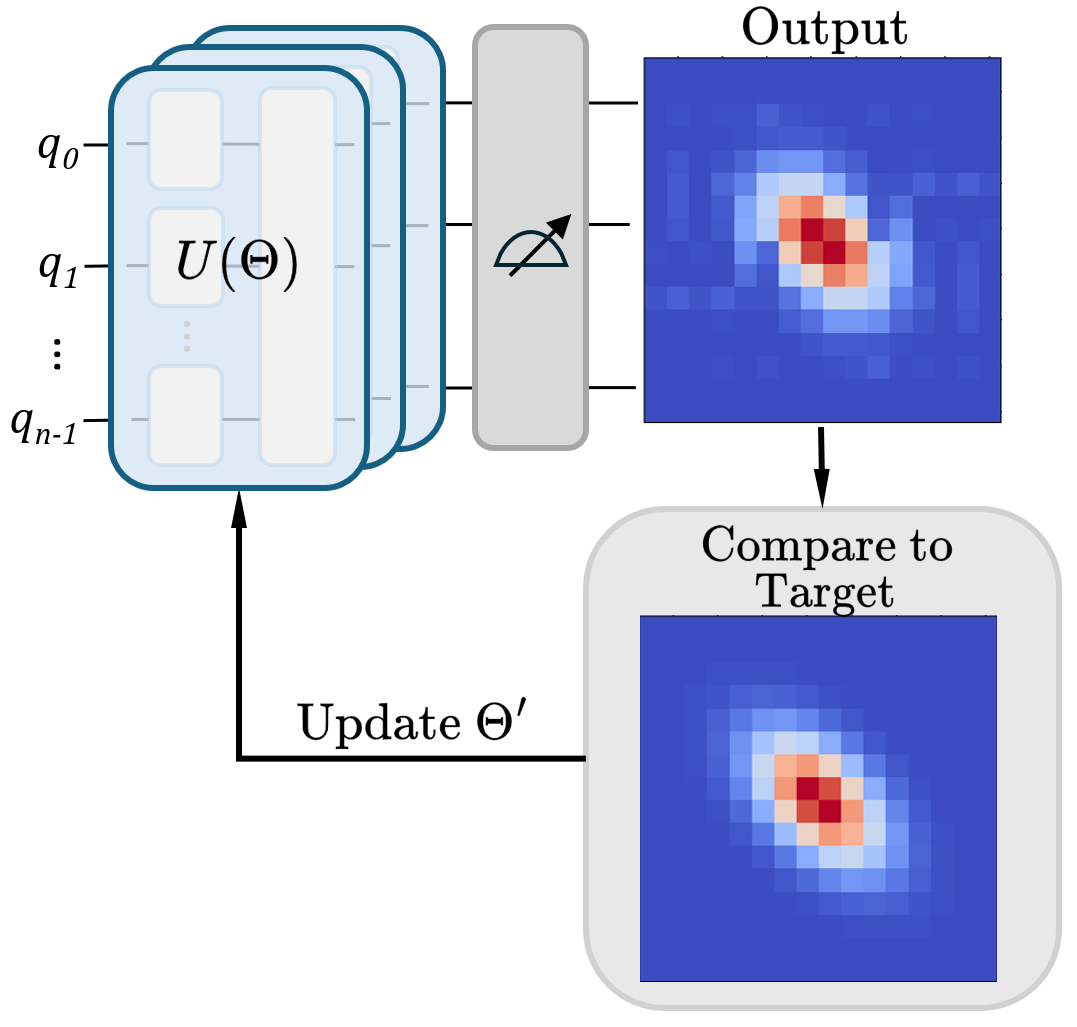}  
\caption{Principle sketch of the training process of the Direct Variational Generator in a quantum variational algorithm. The parameter set ${\bm \Theta}$ of the DVG generates an output distribution, which is compared to the target distribution in the cost function. The generator is rerun iteratively with an updated parameter set ${\bm \Theta}^{\prime}$ until a minimum of the cost function is obtained.}
\label{fig:DVG_Training}
\end{figure}

\subsection{Classical part -- Optimization loop}
\label{sec:Optimization loop}
In the training process, the $n\times d$ parameters $\theta^{(j)}_i$ are adapted such that the target quantum state can be reproduced, $|\Psi\rangle_{\rm gen}\approx |\Psi^{\ast}\rangle$. The optimized set of rotation angles is then used in the DVG after training for the sampling of the fluctuations. As we consider here two-dimensional joint PDFs, the output of the DVG has to be reshaped into a two-dimensional array $\hat{p}$ which is a discretized approximation of the original PDF of velocity fluctuations. 

For the present flow, $p(\xi_1,\xi_2,\xi_3)$ is split into a two-dimensional joint  $p(\xi_1,\xi_3)$ and $p(\xi_2)$. Both problems can be treated separately (as already explained). The column vector ${\bm P} \in \mathbb{R}^{2^n}$, which contains all the probabilities $|a_i|^2$, is then mapped to a two-dimensional matrix $\hat p\in \mathbb{R}^{N \times N}$ with components $p_{ik}$ and $N=2^{n/2}$. For $n=8$, we would have 256 probability amplitudes and can convert $p(\xi_1, \xi_3)$ into an $16\times 16$ array with entries $p_{ik}$. This encoding is possible due to the tensor product structure of the underlying Hilbert space ${\cal H}$ with its exponentially growing dimension. 
\begin{figure}
\centering
\includegraphics[width=0.5\textwidth]{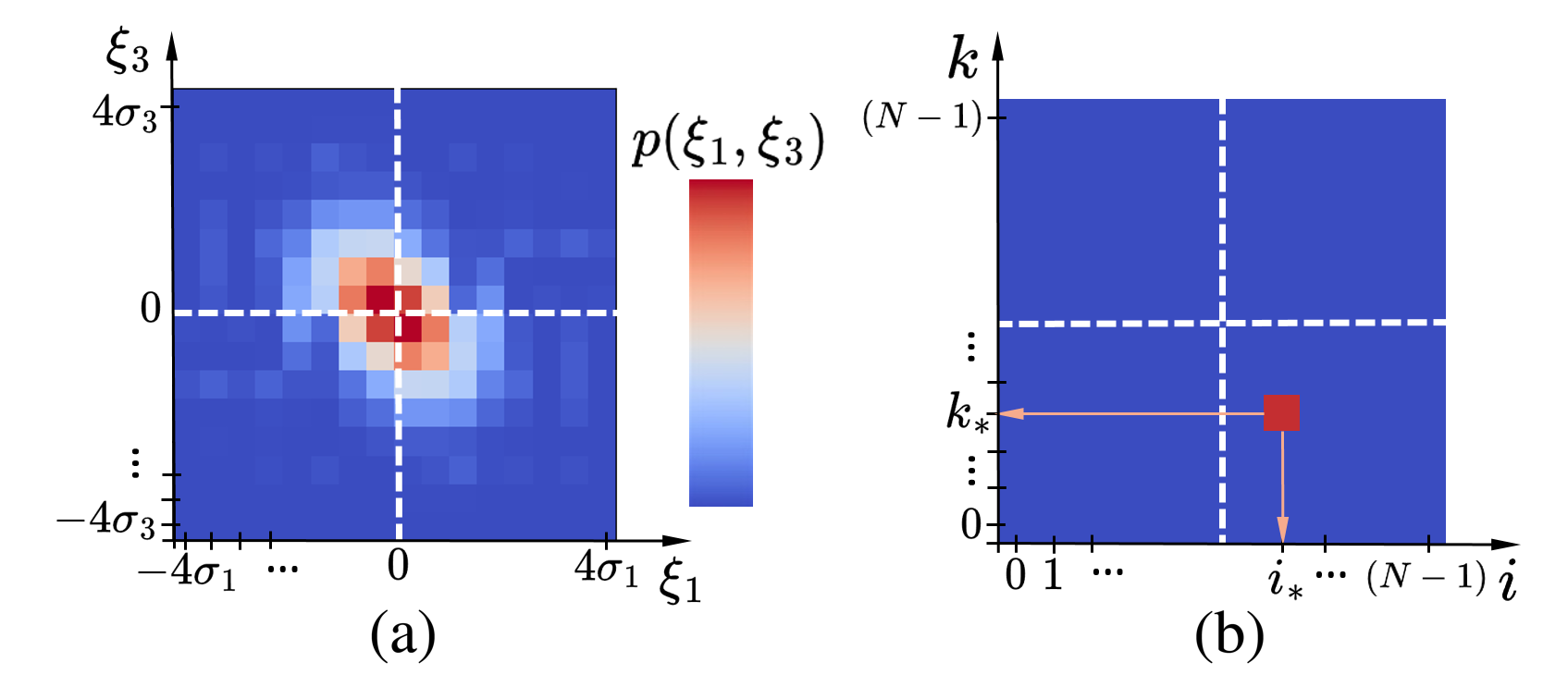}   
\caption{A sketch of the Direct Variational Generator output obtained with (a) an ideal simulation (with a very large number of repeated measurements) and (b) an one-shot measurement for a sampling process. In the two-dimensional case, the range of values of the velocity fluctuations is binned (or discretized) into $N=2^{n/2}$ grid cells. Here, $i^*, k^*$ denote the index of such a particular bin.}
\label{fig:DVG_Measurement}
\end{figure}

In order to obtain the mismatch between the generated output of the DVG and the target, the following standard cost function is evaluated
\begin{align}
    \label{eq:cost function}
    {\rm Cost}({\bm \Theta}) = \|\hat{p}^{\,\rm gen}({\bm \Theta}) - \hat{p}^{\,\rm tar}\|_F^2\,,
\end{align}
where $\hat{p}^{\,\rm gen}$ is the joint PDF generated by the DVG and $\hat{p}^{\,\rm tar}$ is the target distribution, either $p(\xi_1, \xi_3)$ or $p(\xi_2)$. The Frobenius norm is taken in eq.~\eqref{eq:cost function}. 

For the update of the parameter set ${\bm \Theta}$, the Nelder-Mead algorithm \cite{Nelder1965} is used. This is a classical optimization algorithm which is actually designed to solve unconstrained optimization problems without any gradient approximation. The algorithm only uses the function values at some points which construct the simplex in the hyperplane. This simplex is transformed by geometric operations in order to find the minimum of the cost function. Since the optimization algorithm is not designed for quantum problems, we used a two-step optimization approach to find the optimal parameter sets ${\bm \Theta}^{\ast}$. For this, we first parameterize with an initial guess, optimize the parameters and use the optimized parameters as a guess for a second optimization loop. As the second step starts a new unprocessed simplex, this method helps to avoid local minima and stagnation effects due to plateaus by the variational cost function. With this method, we ensure that a good parameter set can be found despite the simple optimization scheme. Furthermore, we applied a mapping in the training that transforms the output into a probability vector with components sorted by magnitude in order to enhance the convergence of the optimization algorithm. In all cases, we use ideal quantum simulations in Qiskit to determine the optimal parameter set. The principle of this training process is shown in Fig.~\ref{fig:DVG_Training}.

\subsection{Sampling Method}
\label{sec:SamplingMethod}
The quantum algorithm aims at sampling the velocity fluctuations with the DVG in correspondence with the PDF that is approximated by $p_{j}$ and $p_{ik}$ for one and two dimensions. For this, a one-shot measurement is used. In a single measurement, the quantum state is projected randomly to one of the $2^n$ eigenstates, e.g., $|l\rangle$. The chosen eigenvector $|l\rangle$ is translated into a grid site $(j)$ and $(i,k)$ that correspond to particular velocity fluctuation magnitudes $\xi_2$ and $(\xi_1, \xi_3)$, as shown in Fig.~\ref{fig:DVG_Measurement} for the two-dimensional case. See also ref. \cite{Benedetti2019} for their data-driven quantum circuit learning approach along similar lines. The one-shot measurement gives ${\bm P}=(0, \dots, 1, \dots 0)^T$ with all components 0 except one at 1 at entry $l_{\ast} \in [0,2^n-1]$. This entry $l_{\ast}$ is mapped to a site $(i_{\ast},k_{\ast})$ of the two-dimensional grid of the size $2^{n/2} \times 2^{n/2}$. The site corresponds to the following velocity fluctuation amplitudes
\begin{align}
    \xi_1 = -4\sigma_1 + i_{\ast} \frac{8 \sigma_1}{N-1} \;\;\mbox{and}\;\;
    \xi_3 = -4\sigma_3 + k_{\ast} \frac{8 \sigma_3}{N-1}
\end{align}
where indices $i_{\ast}$ and $k_{\ast}$ are the column and row indices, respectively.  For the third velocity component, we additionally consider $\xi_2$ from the one-dimensional $p(\xi_2)$ which follows in the same way from $j_{\ast}$. The three velocity fluctuation components $\xi_i$ vary between the corresponding $\pm 4 \sigma_i$, see again Fig.~\ref{fig:DVG_Measurement}.  

\begin{figure}
\centering
\includegraphics[width=0.5\textwidth]{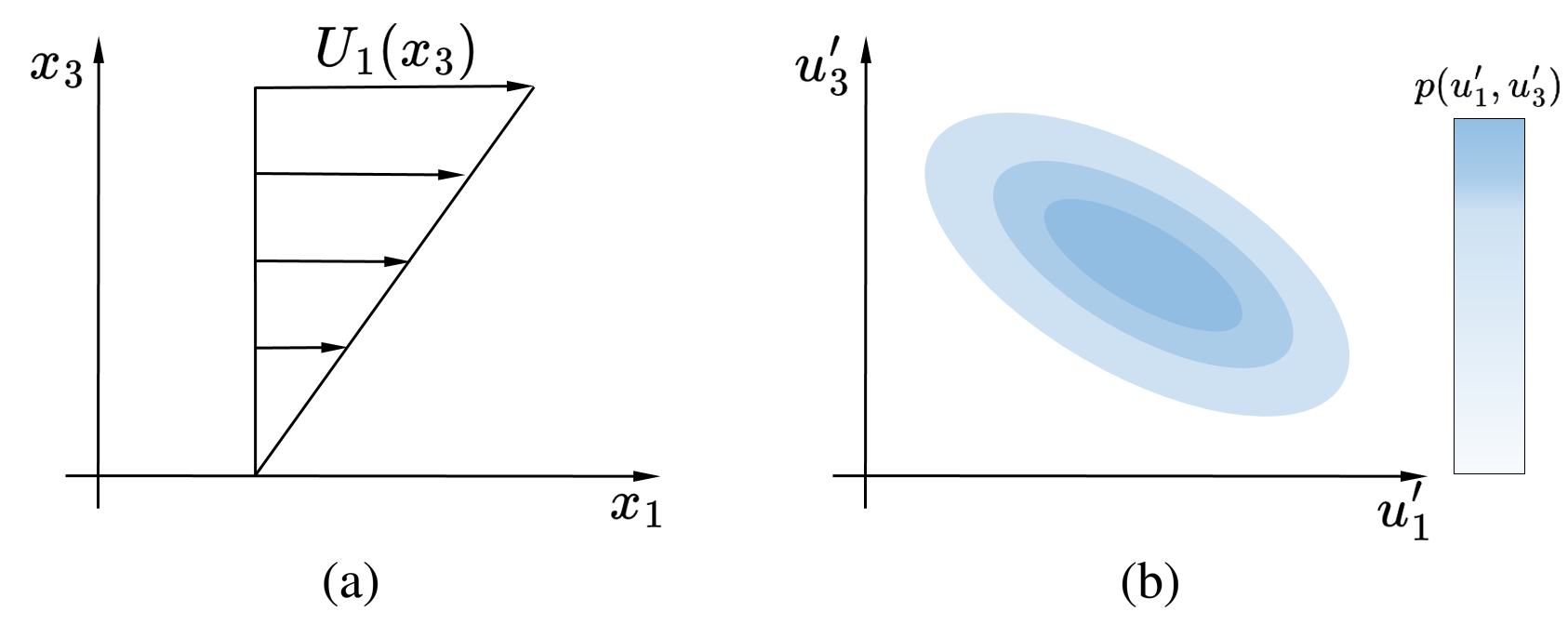}   
\caption{A sketch of (a) the uni-directional mean velocity field $ {\bm U}  (x_3)= U_1(x_3) {\bm e}_1$. The streamwise, $x_1$, and shear directions, $x_3$, form the shear flow plane; the spanwise direction, $x_2$ points into picture plane. (b) The joint probability distribution function of the correlated fluctuations and Gaussian distributed $u_1^\prime$ and $u_3^\prime$.}
\label{fig:shear_flow}
\end{figure}

\section{Results}
\subsection{Lagrangian tracer transport model}
The turbulent shear flow, which we consider for the present study, is composed of a uni-directional mean flow profile, $ U_1(x_3)$ and three-dimensional turbulent velocity fluctuation fields, ${\bm u}^{\prime}({\bm x},t)$, about the mean flow \cite{Pope2000}, as denoted in eq.~\eqref{eq:Lag2}. The HSF\cite{Rogallo1981,Rogers1986,Pumir1996,Schumacher2000,Gualtieri2002,Schumacher2004} is given by
\begin{align}
    {\bm u} (x_1,x_2,x_3,t) = \left[\begin{array}{c} Sx_3 \\ 0 \\ 0 \\\end{array}\right] + 
    \left[\begin{array}{c} u_1^{\prime}(x_1,x_2,x_3,t) \\ u_2^{\prime} (x_1,x_2,x_3,t)\\ u_3^{\prime} (x_1,x_2,x_3,t)\\ \end{array}\right]\,,
\end{align}
where $S$ is the constant shear rate. Figure~\ref{fig:shear_flow}(a) provides a qualitative representation of the shear flow. HSF does not have an outer scale $L$ at which the large-scale shear with rate $S$ is imposed. This idealization was identified as one reason for a strong level of velocity fluctuations in a statistically steady regime in this flow.\cite{Yakhot2003}   

As the numerical integration scheme for the Lagrangian tracers, we employ the Euler–Maruyama method\cite{Higham2021} since the velocity field contains stochastic noise, see also eq.~\eqref{eq:Lag1}. This leads to 
\begin{align}
x_i(t+dt) &= x_i(t) + [U_i+u_i'(t)] \, dt\,,\\ 
u_i^\prime(t+dt) &= u_i^\prime(t) - \frac{u_i^\prime(t)}{\tau_i} dt + \xi_i.
\label{eq:SDE_u}
\end{align}
where $\tau_i$ denotes the Lagrangian integral time, and $\xi_i$ represents a stochastic noise term, which will be modeled by the quantum algorithm. Due to the presence of shear, the fluctuating velocity components $u_1^\prime$ and $u_3^\prime$ become statistically correlated, which arises from correlations in the corresponding noise terms $\xi_i$. The distribution of these stochastic terms is visualized in Fig.~\ref{fig:shear_flow}(b). A detailed description of the stochastic equations and flow characteristics is presented in Appendices \ref{sec:Langevin equ} and \ref{sec:Flow characteristics}, respectively.

The generic procedure in fluid mechanics is to express all scales and fields in units of characteristic physical quantities. Characteristic times in a homogeneous shear flow are measured in units of the shear time, the (viscous) shear length, and the (viscous) shear velocity, which are given by $T_S=S^{-1}$, $ L_S=\sqrt{\nu/S}$, and  $U_S=L_S/T_S=\sqrt{\nu S}$, respectively. Here, $\nu$ is the kinematic viscosity of the flow. These are the definitions for the characteristic quantities at the small-scale end of the shear turbulence, where viscous effects dominate. Alternatively, one can define a characteristic length which is connected to the inertial range of turbulence, the mean kinetic energy dissipation rate $\langle \epsilon\rangle$. This quantity carries the physical dimension of m$^2$s$^{-3}$. The corresponding alternative characteristic units are given by 
\begin{align}
T_S=S^{-1}\,,\;\;  L_S=\sqrt{\frac{\langle\epsilon\rangle}{S^3}}\;\;\mbox{and}\;\;U_S= \sqrt{\frac{\langle\epsilon\rangle}{S}}\,, 
\label{eq:units2}
\end{align}
see e.g. ref. \cite{Gualtieri2002}. These characteristic scales are used in this work to re-scale all quantities. Hence, the dimensionless quantities are given by
\begin{align}
    \label{eq:dimensionless_quants}
    \Tilde{\bm x}= \frac{\bm x}{L_S}, \;\Tilde{{\bm u}}= \frac{\bm u}{U_S}\;\;\text{ and }\;\; \Tilde{t}= \frac{t}{T_S} = S t.
\end{align}
The combination of physical quantities $S$, $\nu$, and $\langle\epsilon\rangle$ allows for a further time scale, the viscous (or Kolmogorov) time scale $\tau_{\eta}=(\nu/\langle\epsilon\rangle)^{1/2}$ \cite{Pope2000}. Consequently, $\tau_{\eta}\ll S^{-1}$ when the shear Reynolds number $Re_S=U_S L_S/\nu=S^{-2}/\tau_{\eta}\gg 1$. These time scales are required when generating synthetic tracer tracks in the turbulent shear flow. 

\subsection{Joint PDF by parametric quantum circuits}
\subsubsection{Statistical correlations in the homogeneous shear flow}
In order to obtain the stochastic noise terms $\xi_i$, we construct a DVG that approximates their joint probability distribution as given in eq.~\eqref{eq:jpdf}. For this, we consider the covariance matrix $\Sigma$ of this distribution derived from the diffusion matrix $B$ as
\begin{align}
\label{eq:Sigma}
    \Sigma = BB^Tdt =  \left[ \begin{array}{ccc}
       \sigma_{1}^2  & 0 & \sigma_{13} \\
        0 &  \sigma_{2}^2 & 0 \\
        \sigma_{13} & 0 & \sigma_{3}^2
    \end{array} \right],
\end{align}
where $\sigma_1^2, \sigma_2^2, \sigma_3^2$ denote the variances of the corresponding stochastic components. A detailed derivation is provided in Appendix~\ref{sec:Langevin equ}. From the off-diagonal element $\sigma_{13}$, we further compute the correlation coefficient between $\xi_1$ and $\xi_3$ as
$\rho_{13} = \sigma_{13}/(\sigma_1 \sigma_3)$. The corresponding joint probability density function (PDF) for the full three-dimensional noise vector ${\bm \xi} = (\xi_1, \xi_2, \xi_3)$ is a multivariate Gaussian distribution and given by
\begin{align}
\label{eq:pdf_3d}
    p({\bm \xi}) = \frac{1}{\sqrt{(2 \pi)^3 \det (\Sigma)}} \exp\left\{ -\frac{1}{2} {\bm \xi}^T \Sigma^{-1} {\bm \xi} \right\}.
\end{align}
Since we only find a correlation for the velocity components in the shear plane and thus for $\xi_1$ and $\xi_3$, we treat $\xi_2$ as independent and factorize the joint distribution accordingly. This gives 
\begin{align}
\label{eq:pdf_1d}
    p(\xi_2) = \frac{1}{\sqrt{2 \pi} \sigma_2} \exp \left\{ \frac{- \xi_2^2}{2\sigma_2^2} \right\}
\end{align} 
for $\xi_2$ and a bivariate normal distribution for $\xi_1$ and $\xi_3$,
\begin{align}
\label{eq:pdf_2d}
    p(\xi_1,\xi_3) &= \frac{1}{2 \pi \sigma_1 \sigma_3\sqrt{ (1-\rho_{13}^2)}} \notag   \\
    &\exp\left\{ -\frac{\sigma_3^2 \xi_1^2 - 2\rho_{13} \sigma_1 \sigma_3 \xi_1 \xi_3 + \sigma_1^2 \xi_3^2}{2\sigma_1^2 \sigma_3^2 (1-\rho_{13}^2)} \right\}\,,  
\end{align}
where $\Sigma^{-1}$ is explicitly written using $\rho_{13}$.

\begin{figure}
\centering
\includegraphics[width=0.5\textwidth]{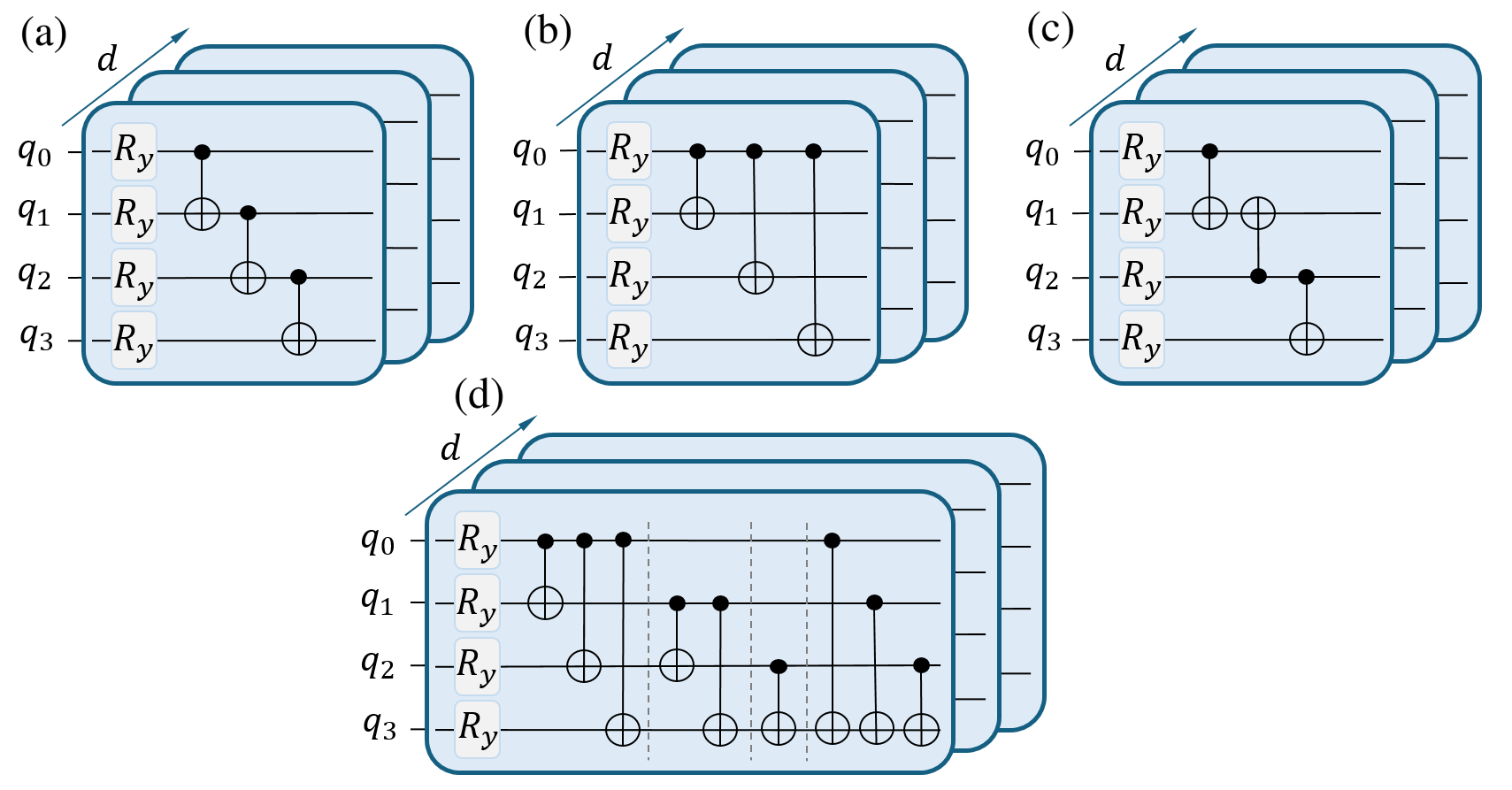}   
\caption{Quantum ansatz structures following ref.~\cite{Nakhl2024} for $n=4$. Here, (a) is the linear circuit, (b) the single control circuit, (c) the alternating circuit and (d) the full circuit and $d$ defines the number of circuit layers in all cases.}
\label{fig:DVG_Ansatz}
\end{figure}

\subsubsection{Quantum Ansatz for Direct Variational Generator}
In this study, we analyze four distinct entanglement structures for the DVG, following the approach proposed in ref.~\cite{Nakhl2024}. Specifically, we consider the (i) linear, (ii) single-control, (iii) alternating, and (iv) full  circuit architectures. The corresponding quantum circuits are illustrated in Fig.~\ref{fig:DVG_Ansatz}. To assess their performance, we evaluate all of them with respect to two benchmark cases, the one-dimensional probability density function (PDF), see \eqref{eq:pdf_1d}, using $n = 5$ qubits, and the two-dimensional PDF, see \eqref{eq:pdf_2d}, using $n = 10$ qubits. These values are selected as the minimum resolution required to capture the relevant distribution features within $2^5=32$ bins for each $\xi_i$. As a measure of performance, we compute the mean squared error (MSE) which is given by eq.~\eqref{eq:cost function} after optimization of parameters for a given circuit architecture.

In our analysis, we vary the number of circuit layers $d$ from 1 to 10 in order to identify an optimal trade-off between accuracy and circuit complexity. Specifically, we aim to determine the best-performing ansatz with the lowest number of quantum gates and circuit layers, ensuring an efficient implementation suitable for near-term quantum hardware. Note that other measures, such as a Kulback-Leibler divergence, would be possible to quantify the difference between both distributions, see e.g. ref. \cite{Koecher2025}, but we focus here on the MSE. 

\begin{figure}
\centering
\includegraphics[width=0.5\textwidth]{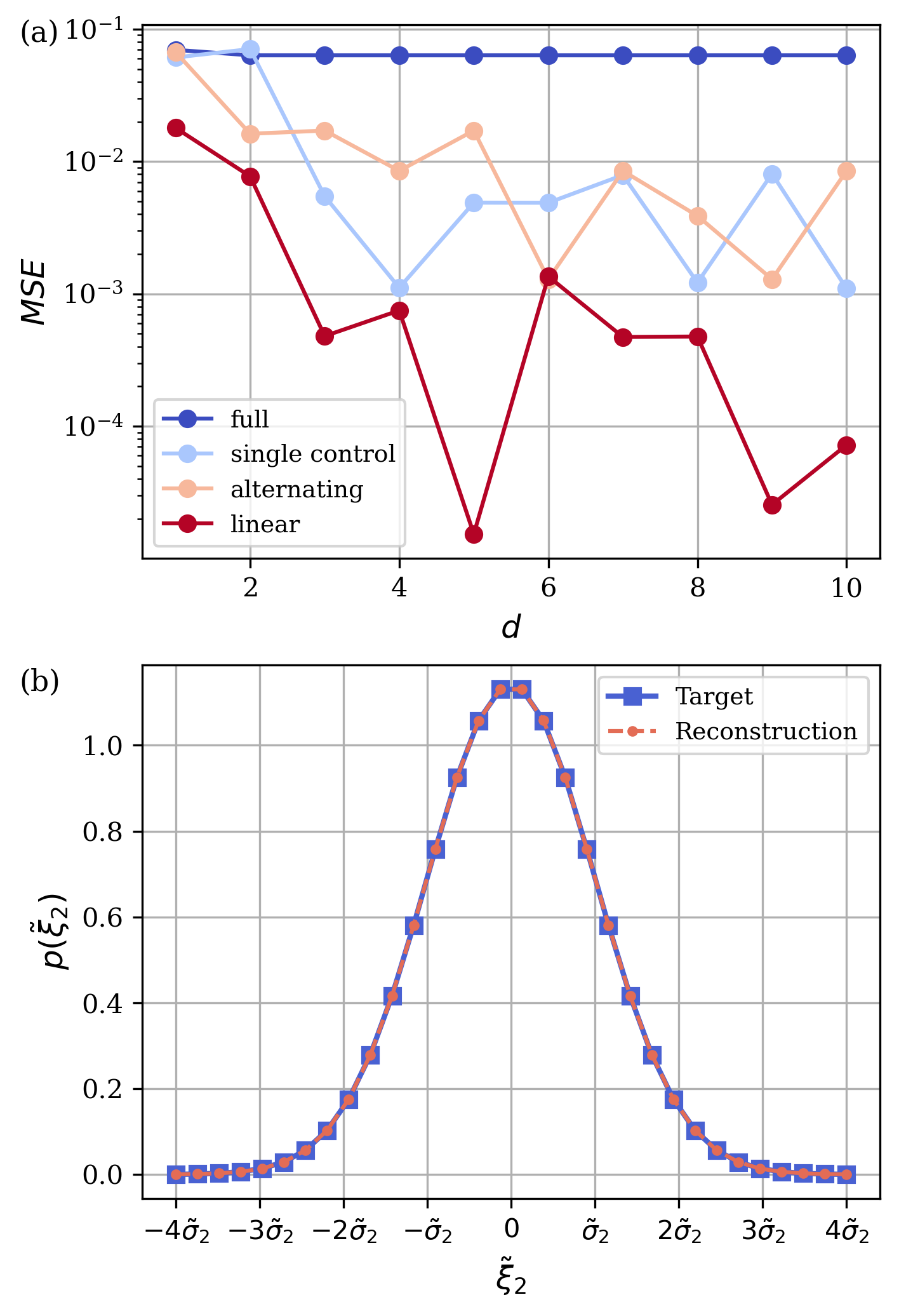}   
\caption{Analysis of the DVG approximating $p(\xi_2)$ where (a) displays the mean squared error (MSE) over the number of circuit layers $d$ and (b) shows the evaluation of the optimal DVG reconstruction with $n=5$ qubits in comparison to the target distribution.}
\label{fig:DVG_Analyse_1d}
\end{figure}
\begin{figure*}
\centering
\includegraphics[width=\textwidth]{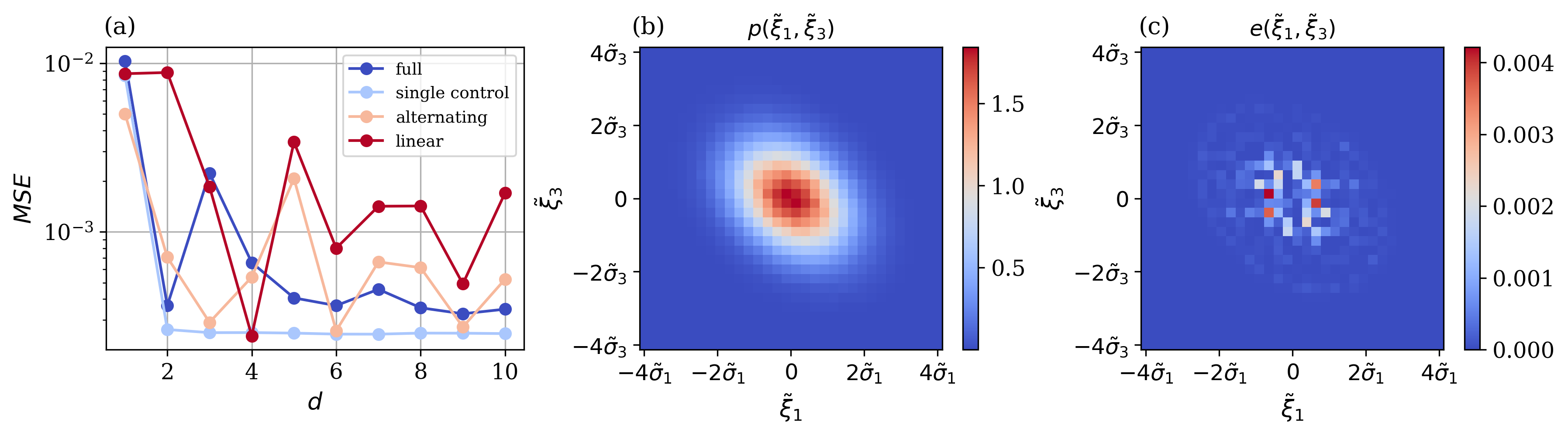}   
\caption{Analysis of the DVG approximating $p(\xi_1,\xi_3)$ with $n=10$ qubits where (a) is the MSE over the number of circuit layers $d$, (b) shows the evaluation of the optimal DVG reconstruction and (c) shows the corresponding squared difference $e$ in comparison to the target PDF.}
\label{fig:DVG_Analyse_2d}
\end{figure*}

In Fig.~\ref{fig:DVG_Analyse_1d}, we present the DVG analysis for the one-dimensional PDF which is given in eq.~\eqref{eq:pdf_1d}. The evaluation of the MSE in Fig.~\ref{fig:DVG_Analyse_1d}(a) indicates that the linear circuit achieves the best overall performance, while the fully entangled ansatz consistently yields the poorest results, with no significant improvement observed even as the number of layers increases. This suggests that the full circuit may not be suitable for approximating this particular target distribution. The single-control and alternating circuits exhibit similar performance trends; the MSE of both falls in between those of the linear and full configurations. In particular, for all circuit architectures, an increase of the number of layers does not necessarily guarantee better approximation accuracy. A possible explanation is that deeper circuits introduce more trainable parameters, which lead to a more complex optimization landscape and avoid better convergence to a global minimum. The optimal configuration in this setting is achieved with a linear circuit using $d=5$ layers. The corresponding reconstructed PDF is shown in Fig.~\ref{fig:DVG_Analyse_1d}(b) in comparison to the target PDF. This configuration is used for all further studies.

Fig.~\ref{fig:DVG_Analyse_2d} shows the DVG analysis for the two-dimensional PDF, see eq.~\eqref{eq:pdf_2d}. The corresponding MSE values for all considered circuit architectures are shown in Fig.~\ref{fig:DVG_Analyse_2d}(a). A significant reduction in the MSE is observed across all ansätze within the first four circuit layers. Beyond this depth, the performance either stops improving, as seen for the single-control circuit, or even gets worse, as in the case of the linear circuit. Based on these results, we select the single-control circuit with $d=3$ layers for further investigations, as it achieves optimal performance with the minimal circuit complexity. The reconstructed two-dimensional PDF and the corresponding pixel-wise squared error relative to the target distribution are depicted in Figs.~\ref{fig:DVG_Analyse_2d}(b) and \ref{fig:DVG_Analyse_2d}(c), respectively.

\begin{figure*}
\centering
\includegraphics[width=\textwidth]{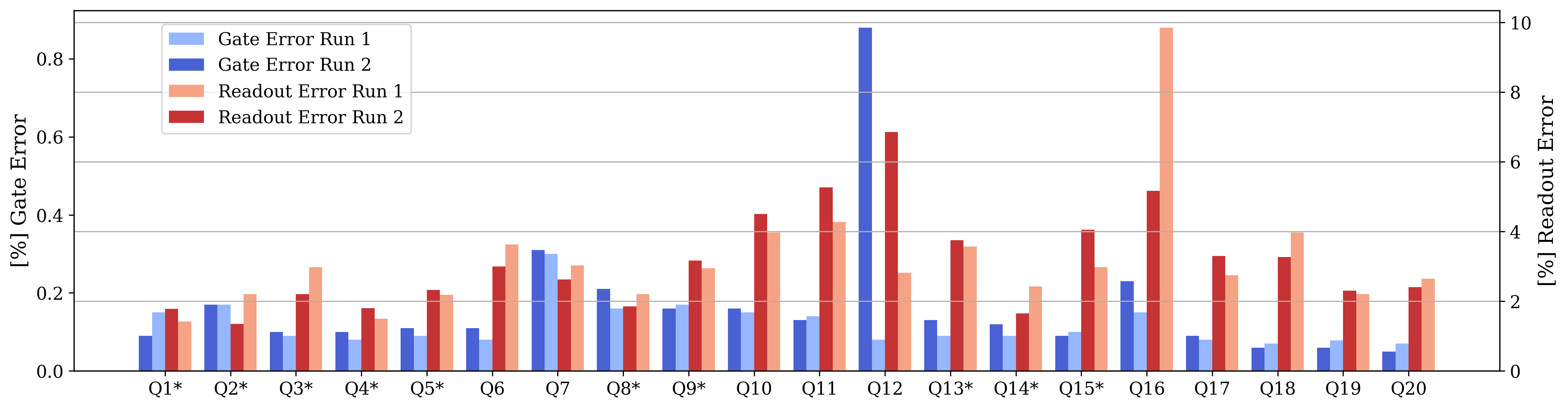}   
\caption{Single qubit gate error and readout error for qubits of IQM Garnet where $*$ marks the qubits chosen for the reconstruction of the PDFs for Runs $1$ and $2$, Run $1$ took place on 05/28/25 and Run $2$ on 06/04/2025.}
\label{fig:Garnet_Errors}
\end{figure*}

\subsection{Reconstruction of the PDFs on IQM Garnet}
In this section we present the reconstruction of the PDFs using the quantum computer IQM Garnet\cite{iqm2024}, a 20-qubit quantum processing unit (QPU). This QPU is based on superconducting transmon qubits and offers high-fidelity quantum operations.~\cite{Marxer2023} The qubits are arranged in a square lattice architecture and connected with tunable couplers. The native gate set includes single-qubit Pauli gates $X$ and $Y$, as well as a two-qubit controlled-$Z$ (C$Z$) gate. Accordingly, all quantum circuits must be transpiled (translated + compiled) to this native gate and connectivity structure. Real QPUs, such as the IQM Garnet, exhibit various sources of noise and errors since the quantum system is never perfectly insulated from its environment, despite cooling down to milli kelvins away from absolute zero temperature. These errors include imperfections in single- and two-qubit gates, such as spontaneous qubit flips, and readout operations. For the reconstruction of the target PDFs, we select the $n=5$ (for the one-dimensional) and $n=10$ (for the two-dimensional) qubits which have the lowest error rates, a quantity that is monitored by the provider. Figure~\ref{fig:Garnet_Errors} illustrates the corresponding single-qubit gate and readout error rates for two representative runs on the QPU on different days. The single-qubit gate errors account for Pauli-type and depolarizing noise during gate application, whereas the readout errors primarily stem from imperfections in the measurement process, often the dominant source of noise and imperfection. For instance, the median gate error on these two days were $0.09\%$ and $0.12\%$ while the median readout errors were $2.89\%$ and $2.81\%$.

\begin{figure*}
\centering
\includegraphics[width=\textwidth]{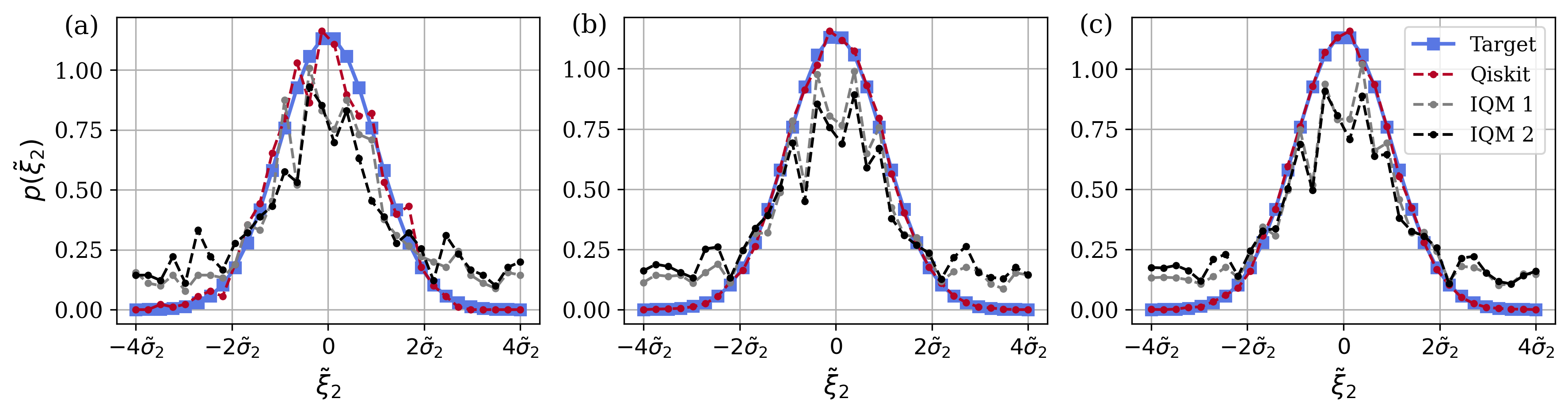}   
\caption{Reconstructed PDFs of run $1$ and $2$ for (a) $1000$, (b) $10000$ and (c) $20000$ shots with $n=5$ on qubits $Q1-Q5$. The result of the Qiskit Statevector Sampler is in red; those of the real IQM Garnet runs in grey and black. All are compared to the target distribution.}
\label{fig:DVG_Reconstruction_1d}
\end{figure*}

\begin{figure*}
\centering
\includegraphics[width=\textwidth]{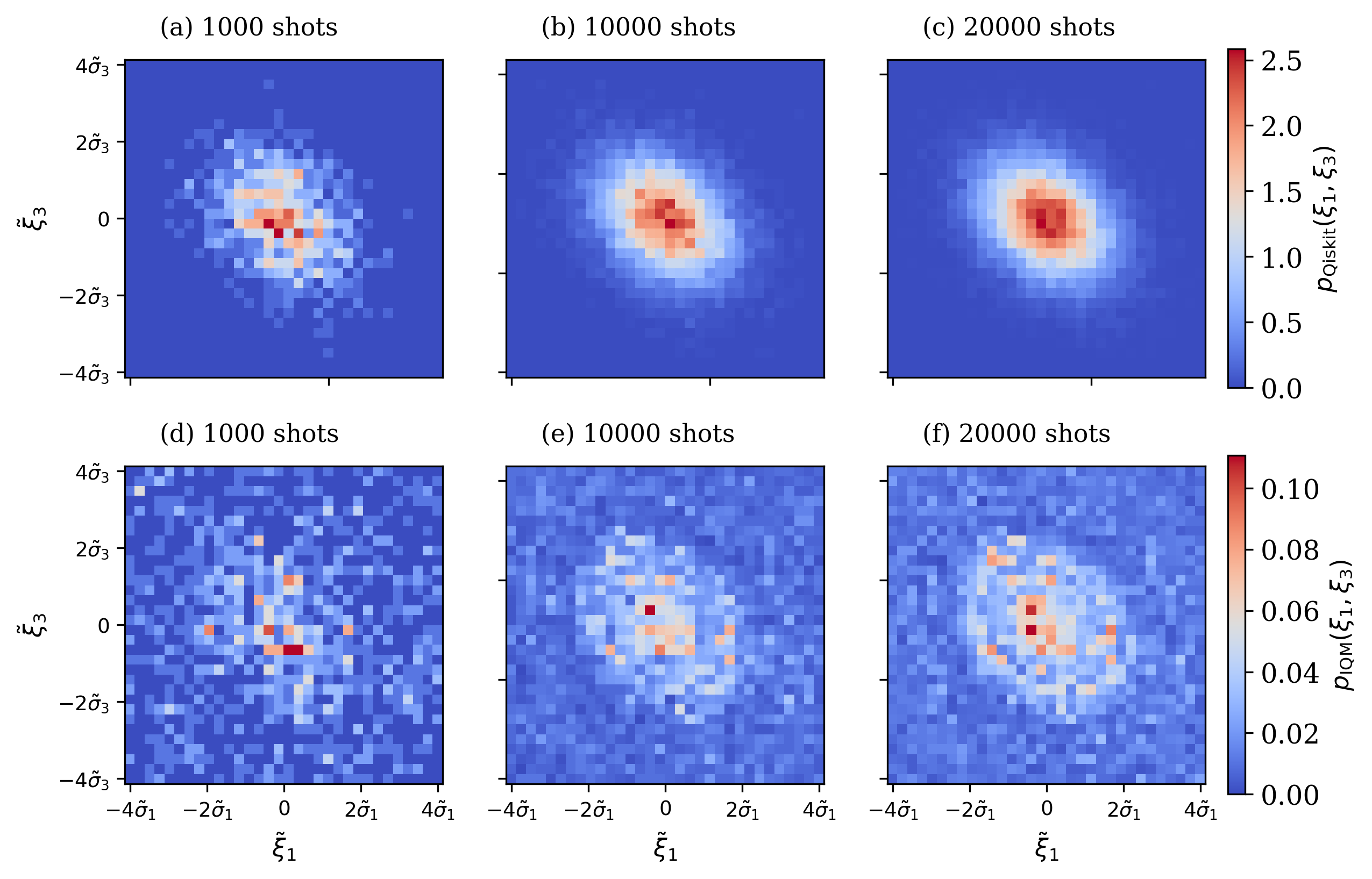}   
\caption{Reconstructed two-dimensionla PDFs with $n=10$ qubits where panels (a)--(c) show the Qiskit Statevector Sampler results and panels (d)--(f) the IQM Garnet results for run $1$.}
\label{fig:DVG_Reconstruction_2d}
\end{figure*}

The reconstructed one-dimensional and two-dimensional PDFs are shown in Figs.~\ref{fig:DVG_Reconstruction_1d} and \ref{fig:DVG_Reconstruction_2d}, respectively. For comparison, we evaluate the DVGs using the Qiskit Statevector Simulator~\cite{Qiskit} next to the IQM Garnet backend. The Qiskit Statevector Simulator is an idealized simulation tool for quantum algorithms, introducing only statistical noise due to sampling, and no physical noise. As expected, the simulator accurately reproduces the peaks and tails of the one- and two-dimensional PDFs, including regions where values are close to zero. Increasing the number of measurement shots further improves the reconstruction quality, which we have verified. In contrast, results obtained from the IQM Garnet backend reflect the influence of physical noise, particularly during readout and gate execution. While the peak of the distribution is qualitatively reconstructed, noise significantly degrades overall fidelity, and increasing the number of shots offers limited improvement only. 

\subsection{Tracer trajectories and particle dispersion}
Now that we have presented the DVG, which is used as a quantum sampling device in subsections III B and C, we proceed with the actual application to the fluid flow problem at hand.  We present the results of the Lagrangian tracer transport in the considered HSF in this subsection. 

To this end, the results of the proposed hybrid classical--quantum method, the quantum-assisted stochastic differential equation (QSDE), are compared with the classical stochastic differential equation model (SDE). This model was developed by Pope \cite{Pope2002} and is detailed in Appendix~\ref{sec:Langevin equ}. In Appendix~\ref{sec:Flow characteristics}, we also provide the concrete parameters of the configuration in table~\ref{tab:Characteristic scales}. This includes for example the shear rate or the mean kinetic energy dissipation rate. They have been obtained from direct numerical simulation (DNS) of an HSF by Sawford and Yeung.\cite{Sawford2000,Sawford2001} In Appendix~\ref{sec:Flow characteristics}, we list in addition the concrete matrices for the correlations of the velocity components in the HSF, the stochastic diffusion, and the resulting Lagrangian integral time scales.  Furthermore, we compare our findings with a standard classical sampling method, a Markov-chain Monte Carlo sampling method (MCMC), which is explained in brief in Appendix~\ref{sec:MCMC}. 

For the QSDE framework, we employ the Qiskit Statevector Sampler for the quantum simulation, since the evaluation on a real QPU leads to noisy reconstructions of the PDFs, and thereby distorting statistical behavior of the particle transport. In Appendix~\ref{sec:1dRandomWalkIQM}, we show a reconstruction of a one-dimensional Gaussian PDF for 6 qubits on IQM first and test the algorithm with a simple one-dimensional Brownian motion $x(t)$ afterwards. The error rates of the gates in the real device lead to a deviation from the expected single particle dispersion, $\langle x^2\rangle=2Dt$ with a given diffusion constant $D$.     

\begin{figure*}
\centering
\includegraphics[width=\textwidth]{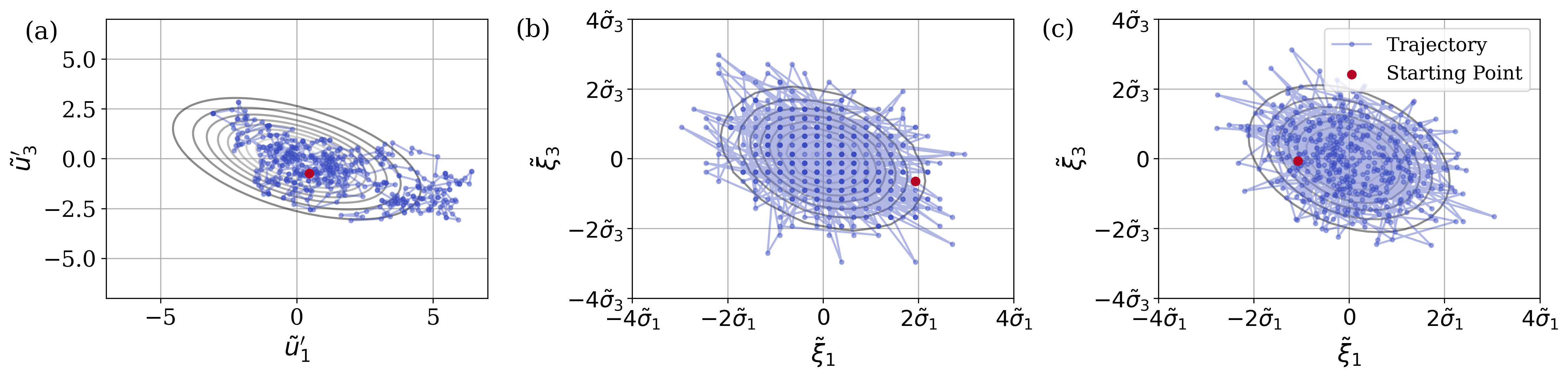}   
\caption{Trajectories of the first $500$ drawn fluctuation velocity samples or noise samples, respectively, for (a) the classical Markov-chain Monte Carlo sampling, (b) the quantum one-shot method and (c) the normal distribution of the classical stochastic differential equation scheme.}
\label{fig:Trajectories}
\end{figure*}

\begin{figure*}
\centering
\includegraphics[width=\textwidth]{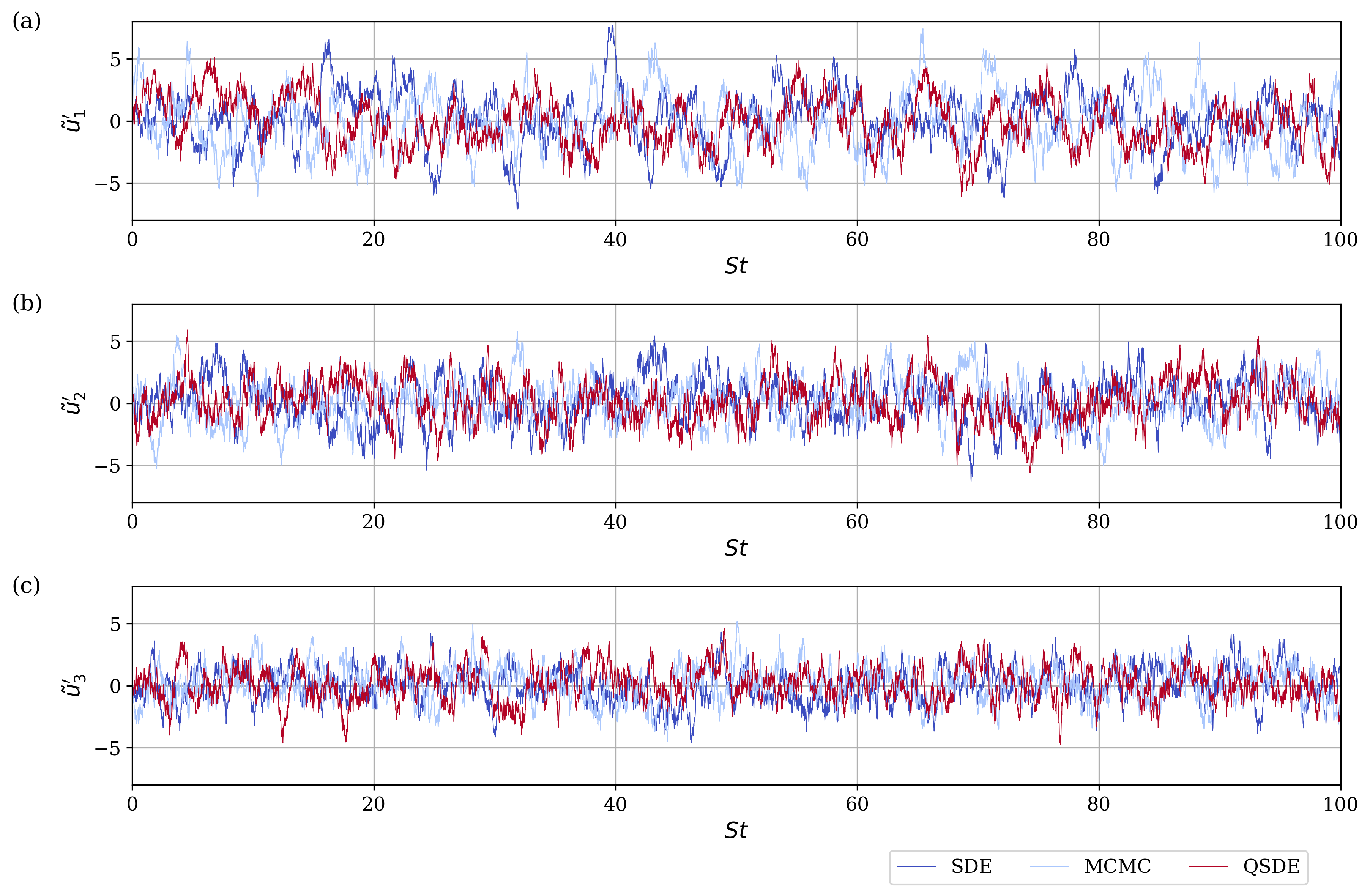}   
\caption{Temporal evolution of the small-scale turbulent velocity components $u_i^{\prime}(t)$ along a specific Lagrangian tracer particle. The classical Markov-chain Monte Carlo scheme (MCMC), the classical stochastic differential equation scheme (SDE) and the hybrid classical-quantum stochastic differential equation scheme (QSDE) are compared for $i=1, 2$ and 3 and panels (a), (b) and (c), respectively.}
\label{fig:Fluctuations}
\end{figure*}

We set the time step $d\tilde{t}=0.01$ and evaluate $N_p = 1000$ Lagrangian tracer paths for the following statistical investigations. Figure~\ref{fig:Trajectories} shows the trajectories of the first $500$ samples. In case of the MCMC, the fluctuation velocities are directly sampled from the joint probability distribution $p(\tilde{u}_1^\prime,\tilde{u}_3^\prime)$ as illustrated in Fig.~\ref{fig:Trajectories}(a). Since each MCMC sample depends on the previous one, the resulting points form a connected trajectory. In contrast, both the QSDE and SDE method approach sample the noise terms $\tilde{\xi}_i$ as shown in Figs.~\ref{fig:Trajectories}(b) and (c). For the QSDE method, the sampled noise values are discretized, reflecting the finite resolution imposed by the small number of qubits. However, the SDE sample points form a denser distribution since they are sampled from an interval. In both methods, QSDE and SDE, the noise samples are statistically independent and randomly distributed from the PDF. The velocity fluctuations are then computed via eq.~\eqref{eq:SDE_u}. The resulting temporal evolution of an individual Lagrangian particle track for all three components of the velocity is presented in Fig.~\ref{fig:Fluctuations}. It is seen that the stochastic example paths, which are obtained from all three methods, agree qualitatively and are in the same range of values. 

\begin{figure}
\centering
\includegraphics[width=0.5\textwidth]{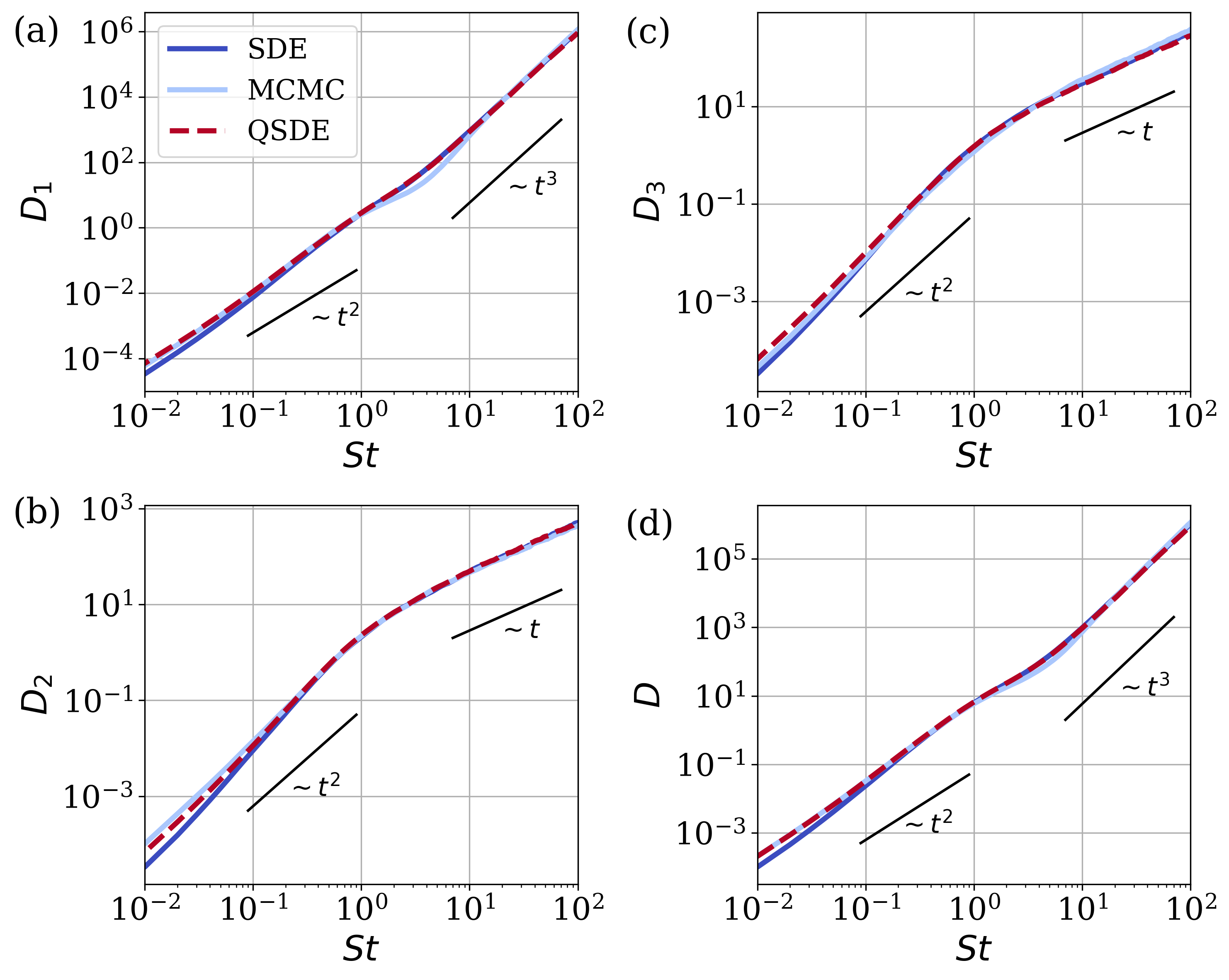}   
\caption{Temporal evolution of the Lagrangian particle pair dispersion. We compare the classical Markov-chain Monte Carlo scheme (MCMC), the classical stochastic differential equation scheme (SDE) and the hybrid classical-quantum stochastic differential equation scheme (QSDE). (a) Pair dispersion along the streamwise direction. (b) Pair dispersion along the spanwise direction. (c) Pair dispersion along the shear direction. (d) Total pair dispersion. Times are given in dimensionless units $St$.}
\label{fig:Dispersion}
\end{figure}

For the statistical analysis, we consider now Lagrangian tracer pairs with a small initial separation. The relative displacement between two individual Lagrangian tracer particles A and B is given by the distance vector
\begin{align}
    {\bm r} = {\bm x}^{\rm A} - {\bm x}^{\rm B}.
\end{align}
The component--wise particle tracer pair dispersion for each spatial direction is computed as
\begin{align}
    D_i(t) = \frac{1}{N_p} \sum_{j=1}^{N_p} [r_{i,j}(t) - r_{i,j}(0)]^2,
\end{align}
where $N_p$ is the number of evaluated paths and $i=1, 2, 3$. The total particle pair dispersion is given by 
\begin{equation}
D(t)=\sum_{i=1}^3 D_i(t)=\langle ({\bm r}(t)-{\bm r}(0))^2\rangle_{N_p}\,. 
\end{equation} 
The corresponding results are shown in Fig.~\ref{fig:Dispersion}. The dispersion curve of proposed hybrid classical--quantum approach (QSDE) shows a very good agreement with the results from both classical approaches, SDE and MCMC. In the directions $i=2,3$, an expected  transition from the ballistic regime ($D_{i}\sim t^2$) to the diffusive regime ($D_{i}\sim t$) is observed around $St\approx 1$. For sufficiently long times, $t\gtrsim S^{-1}$, the Lagrangian correlations between the tracers within the pairs are decayed; their relative motion becomes a Brownian walk. This is also observed in DNS of three-dimensional homogeneous shear turbulence, see the results of Shen and Yeung \cite{Shen1997}. In the streamwise direction, the particle pair dispersion $D_1$ is dominated by the imposed mean shear flow $U_1(x_3)$ in this direction. Here, we observe a transition from $D_1\sim t^2$ to $D_1\sim t^3$, which would actually correspond to a Richardson-like scaling in the inertial range of homogeneous isotropic turbulence, see e.g. Boffetta and Sokolov\cite{Boffetta2002} for two-dimensional turbulence or Biferale et al.\cite{Biferale2005} for the three-dimensional case. It is actually also seen in the HSF simulations by Shen and Yeung.\cite{Shen1997} Panel (d) of the figure shows that the $D_1$ term is the dominant one for $D(t)$ since it gives the same $t^3$-scaling of the two-particle dispersion. We can conclude that the quantum-assisted Lagrangian tracer model provides the same dispersion curves as the classical cases. We have also verified that the single particle dispersion agrees, which is not shown here.  

\section{Discussion and Outlook}
The aim of this work is to present a proof-of-concept investigation of a quantum-assisted algorithm that models turbulent Lagrangian particle transport, in particular the tracer pair dispersion, in a turbulent shear flow. We consider the simplest turbulent shear flow, a homogeneous shear flow, which is characterized by a constant shear rate $S$ along the $x_3$ direction and a steady uni-directional mean flow. In this model flow, a shear stress, i.e., a correlation between the streamwise and shear direction (or wall-normal otherwise) velocity components, $u_1^\prime$ and $u_3^\prime$, is induced. All velocity components have a Gaussian statistics. The tracer motion is thus modeled classically by a system of coupled stochastic differential equations.\cite{Pope2002} Velocities can be sampled from a joint probability density function $p(\xi_1,\xi_3)$. The remaining spanwise velocity component $u_2^{\prime}$ is uncorrelated to the other two components and thus sampled from the PDF $p(\xi_2)$. 

In the proposed hybrid classical--quantum approach, the stochastic model is solved classically and the sampling process of the noise terms is performed by a one-shot measurement in a quantum algorithm. To this end, the PDFs $p(\xi_1,\xi_3)$ and $p(\xi_2)$ are reconstructed with a direct variational generator (DVG) with $n=5$ qubits (which corresponds to 32 bin that intersect the range of values). This generative quantum algorithm involves an optimization process to identify the most efficient quantum circuit architecture with optimized parameters $\Theta$ and circuit depth $d$. We used ideal statevector simulations in combination with the Nelder–Mead optimization method to identify effective circuits for the PDF reconstruction. 

The best performing circuit architecture of our study is the linear ansatz as shown in Fig.~\ref{fig:DVG_Ansatz}(a). An optimal layer depth was $d=5$ for the reconstruction of $p(\xi_2)$. A different ansatz, the single control ansatz with $d=3$ layers, turned out to be the optimum for $p(\xi_1,\xi_3)$. This underlines that the optimal circuits architectures in variational quantum algorithms are often problem-specific. General rules, besides the creation of a sufficient degree of entanglement, are difficult to extract. 

The reconstruction was evaluated using both, an ideal Qiskit statevector simulator (accounting for statistical noise only) and the 20-qubit IQM Garnet QPU (including both statistical and physical noise). Our results demonstrated that current noise levels on real NISQ hardware still limit an accurate PDF reconstruction. An error mitigation could be obtained by including additional qubits (typically 100 times as many physical qubits in a surface code) for an error correction. This goes beyond the scope of this work and the current capabilities of the machine. Therefore, the evaluation of Lagrangian particle transport is based on ideal quantum simulations performed in Qiskit.\cite{qiskit2024}

For each time step, we employ a one-shot measurement strategy to sample the noise terms. The quantum simulator is thus a sampling device. The particle trajectories are then computed classically using the Euler–Maruyama scheme. An analysis of particle pair dispersion demonstrates that the quantum-assisted algorithm yields results that are consistent with classical approaches. As a reference, we compared our findings with two classical methods, direct numerical integration of the SDE model and a classical Markov-chain Monte Carlo sampling method.

The present application case builds on simple Gaussian velocity statistics and thus it is to our view appropriate for a proof-of-concept study (as already stated above). The potential near-term future application that we have in mind for the presented framework is that of a subtask on a quantum computer, namely modeling the turbulent tracer transport in a complex three-dimensional shear flow. The latter would still be simulated on a classical supercomputer.  Already in a turbulent channel flow or a flat-plate boundary layer -- two possible extensions of the homogeneous shear flow to realistic configurations -- the velocity fluctuations will become height-dependent and have a non-Gaussian statistics when approaching the boundary or wall.\cite{Schlatter2010} On the one hand, this will complicate the construction of the multi-dimensional sampling distribution which is not anymore a multi-variate Gaussian. On the other hand, it can be expected that particularly in such an application case quantum sampling reaches an advantage over classical sampling methods, by making use of the tensor product structure of the underlying multi-qubit Hilbert space, which grows exponentially with the number of qubits. 

Future work on this subject will also involve different ways to model tracer dispersion: this can include (1) a complete treatment of the Lagrangian tracer module on a quantum device including the solution of the ordinary differential equations, (2) a modified quantum walk, the analogue of Brownian walks on quantum computers, or (3) a recently developed quantum MCMC algorithm.\cite{Layden2023} These studies are currently in progress and will be reported elsewhere.   

\section*{Data Availability Statement}
The data and programs that support the findings of this study will be made available once the manuscript is accepted for publication.

\acknowledgements
The research of J.I. is funded by the European Union (ERC, MesoComp, 101052786). Views and opinions expressed are however those of the authors only and do not necessarily reflect those of the European Union or the European Research Council. We acknowledge the use of the 20-qubit IQM Garnet for this study and thank for their technical support. The views expressed are those of the authors.  F.S. is supported by the Deutsche Forschungsgemeinschaft. We thank Marco Dentz and Philipp Pfeffer for helpful discussions and suggestions.

\appendix
\section{Primer on quantum algorithms}
\label{sec:introQC}
We present some basic definitions of quantum computing and refer to Nielsen and Chuang \cite{Nielsen2010} for more details on the fundamentals. While a single classical bit can take two discrete values, namely $\{0,1\}$ only, a single quantum bit (in short qubit) is a superposition of two basis states in the vector space $\mathbb{C}^2$. The quibit can take any state, i.e., infinitely many, on the surface of a unit sphere, which is termed the Bloch sphere,
\begin{equation}
|q_1\rangle=c_1|0\rangle+c_2|1\rangle=c_1
\left(
\begin{array}{c}
1 \\ 0\\
\end{array}\right)
+c_2
\left(
\begin{array}{c}
0 \\ 1\\
\end{array}\right) \in \mathbb{C}^2\,, \label{Eq:B1}
\end{equation}
with $c_1, c_2\in \mathbb{C}$ and $\sqrt{|c_1|^2+|c_2|^2}=1$ . Vectors $|0\rangle$ and $|1\rangle$ are the basis vectors in the so-called Dirac notation. \cite{Nielsen2010} A two-qubit state vector, which is 4-dimensional, is obtained as the tensor product of two single-qubit vectors, 
\begin{equation}
|\Psi\rangle=|q_1\rangle\otimes |q_2\rangle\,.
\end{equation}
The basis of this tensor product space $\mathbb{C}^2\otimes\mathbb{C}^2$ is consequently given by 4 vectors: $|j_1\rangle=|0\rangle\otimes |0\rangle$, $|j_2\rangle=|0\rangle\otimes |1\rangle$, $|j_3\rangle=|1\rangle\otimes |0\rangle$, and $|j_4\rangle=|1\rangle\otimes |1\rangle$. This can be generalized to $n$ qubits. An $n$-qubit quantum state, which is given by
\begin{equation}
   |\Psi\rangle = \sum\limits_{k=1}^{2^n} c_k |j_k\rangle \hspace*{1em} \text{with} \hspace*{1em} \sum\limits_{k=1}^{2^n} |c_k|^2 = 1\,,
\end{equation}
is termed fully separable if it can be written as
\begin{equation}
   |\Psi\rangle = \overset{n}{\underset{i=1}{\bigotimes}} \,|q_i\rangle \,,
\end{equation}
where $|q_i\rangle$ are single qubit quantum states given by eq.~\eqref{Eq:B1}. Not fully separable quantum states are called entangled. 

The evolution of a quantum state (with respect to time $t$) corresponds to a linear unitary transformation, $|\Psi(t)\rangle=U(t)|\Psi(0)\rangle$ with $U^{-1}=U^{\dagger}$. These transformations are performed by elementary single- and two-qubit gates, which correspond to complex $2\times 2$ and $4\times 4$ matrices, respectively. Such matrices are the Pauli matrices $X, Y, Z$ or the rotation matrices $R_X(\theta)=\exp(-i\frac{\theta}{2} X)$, $R_Y(\theta)=\exp(-i\frac{\theta}{2} Y)$, and $R_Z(\theta)=\exp(-i\frac{\theta}{2} Z)$. Quantum circuits are made up of several single- and two-qubit gates which operate on multiple qubits. A quantum state is obtained by a measurement.\cite{Nielsen2010}

\section{Langevin equations of tracer dispersion}
\label{sec:Langevin equ}
The motion of Lagrangian tracers in a turbulent flow can be described in a stochastic model as a superposition of drift and diffusion, starting with G. I. Taylor's seminal work \cite{Taylor1921}, see also refs. \cite{Haworth1986,Thomson1987,Wilson1993,Pope2002} for more recent formulations and extensions. Starting point is the following system of equations that describes the evolution of a spatial and velocity increment of a particular tracer particle from time $t$ to $t+dt$ given by
\begin{align}
dx_i &= [ U_i +u_i^{\prime}]\, dt\,, \label{eq:sde1a} \\
du_i^{\prime} &= - A_{ij} u^{\prime}_j dt + \underbrace{B_{ij} dW_j}_{=\xi_i} \label{eq:sde1b}\,.
\end{align}
Eq.~\eqref{eq:sde1a} is deterministic while eq.~\eqref{eq:sde1b} is stochastic due to the vector-valued Wiener process $W_j(t)$ for $j=1,2,3$ on the right hand side of \eqref{eq:sde1b} with
\begin{align}
\langle dW_i\rangle =0 \quad \mbox{and} \quad \langle dW_i dW_j\rangle = \delta_{ij} dt  \label{eq:sde1c}\,. 
\end{align}
This system of equations generates stochastic tracer trajectories which are used to quantify the turbulent dispersion. The velocity fluctuation components are Gaussian distributed and given by a multivariate PDF $p(u_1^{\prime},u_2^{\prime},u_3^{\prime})$. The components of the covariance matrix $\hat{C}$ for the present shear flow cases follow to $C_{ij}=\text{Cov}(u_i^{\prime},u_j^{\prime}) = \langle u_i^{\prime} u_j^{\prime}\rangle - \langle u_i^{\prime} \rangle \langle u_j^{\prime} \rangle$. With $\langle u_i^{\prime}\rangle=0$ this results to
\begin{align}
        \hat C = \left[ \begin{array}{ccc}
       \langle u_1^{\prime\,2} \rangle & 0 & \langle u_1^{\prime} u_3^{\prime} \rangle \\
       0 & \langle u_2^{\prime\, 2} \rangle & 0  \\
       \langle u_1^{\prime} u_3^{\prime} \rangle & 0 & \langle u_3^{\prime\,2} \rangle
    \end{array}\right]\,.
\label{eq:cov}    
\end{align}
The correlation $u_1^{\prime}$ and $u_3^{\prime}$, the Reynolds shear stress, can be rewritten to the correlation coefficient $\rho$ which is given by 
\begin{align}
    \rho_{13} = \frac{\langle u_1^{\prime} u_3^{\prime} \rangle}{\sqrt{\langle u_1^{\prime\,2} \rangle \langle u_3^{\prime\,2} \rangle}}\,.
\end{align} 
The correlation coefficient can take values in $\rho \in (-1,1)$ where $\rho=0$ implies that no correlation is present while $\rho\to 1, -1$ implies highest degree of correlation. A higher shear rate $S$ leads to a stronger correlation $\rho \propto S$. From eq.~\eqref{eq:sde1b} follows an evolution equation for the covariance matrix which is given by \cite{Pope2002}
\begin{align}
\frac{d C_{ij}}{dt} = - A_{ik} C_{kj}-A_{ik} C_{kj}^T + B_{ik}B_{kj}^T \,,
\label{eq:cov1}    
\end{align}
We used \eqref{eq:sde1c} to obtain
\begin{align}
\langle B_{ik} B_{jm} dW_k dW_m\rangle &= B_{ik} B_{jm} \langle dW_k dW_m\rangle \nonumber\\
   &= B_{ik} B_{jm} \delta_{km} dt \nonumber \\ 
   &= B_{ik} B_{kj}^T dt\nonumber\,,
\end{align}
after averaging.
Following ref. \cite{Thomson1987}, it is assumed that $p(u_1^{\prime},u_2^{\prime},u_3^{\prime})$ is steady. The matrix $A_{ij}$ in \eqref{eq:sde1b} follows to 
\begin{align}
    A_{ij} = \frac{\delta_{ij}}{\tau_i} \quad\mbox{with}\quad \tau_i=\int_0^{\infty}\langle u_i^{\prime}(t) u_i^{\prime}(t+s)\rangle_t\, ds\,.
\label{eq:tau}    
\end{align} 
Here, $\tau_i$ are the Lagrangian integral times which quantify the short-term memory of the velocity fluctuations $u_i^{\prime}$. They will be considered as model parameters in our approach. Statistical stationarity and \eqref{eq:tau} imply that eq.~\eqref{eq:cov1} simplifies to
\begin{align}
C_{ij}=\frac{\tau_i}{2} B_{ik} B_{kj}^T\,.
\label{eq:cov3}
\end{align}
Thus we get 
\begin{align}
du_1^{\prime} &= - \frac{u^{\prime}_1}{\tau_1} dt + B_{11} dW_1 + B_{13} dW_3\label{eq:sde2a}\,,\\
du_2^{\prime} &= - \frac{u^{\prime}_2}{\tau_2} dt + B_{22} dW_2 \label{eq:sde2b}\,,\\
du_3^{\prime} &= - \frac{u^{\prime}_3}{\tau_3} dt + B_{33} dW_3 + B_{31} dW_1\label{eq:sde2c}\,,
\end{align}
where symmetry of the diffusion tensor is assumed, $B_{13}=B_{31}$. We modify the covariance matrix \eqref{eq:cov} to 
\begin{align}
        \hat{\tilde C} = \left[ \begin{array}{ccc}
       \dfrac{\langle u_1^{\prime\,2} \rangle}{\tau_1} & 0 & \dfrac{\langle u_1^{\prime} u_3^{\prime} \rangle}{2\tau_{13}} \\
       0 & \dfrac{\langle u_2^{\prime\, 2} \rangle}{\tau_2} & 0  \\
       \dfrac{\langle u_1^{\prime} u_3^{\prime} \rangle}{2\tau_{13}} & 0 & \dfrac{\langle u_3^{\prime\,2} \rangle}{\tau_3}
    \end{array}\right]\,,
\end{align}
with timescale $\tau_{13}^{-1}=\tau_1^{-1}+\tau_3^{-1}$, see ref. \cite{Wilson1993}. The modified covariance matrix is connected to the diffusion tensor by
\begin{align}
        \tilde C_{ij} = \frac{1}{2} B_{ik} B_{jk}\,.
\end{align}
Since $\tilde C_{ij}$ is still symmetric, it can be diagonalized by means of an orthogonal matrix $\hat O$. This results $\tilde{C}_{ij}=O_{im}\Lambda_{mk}O^T_{kj}$ where $\Lambda_{mk}=\lambda_m\delta_{mk}$ is a diagonal matrix. When all eigenvalues $\lambda_m>0$, we can take the square root of $\Lambda$. Thus the diffusion matrix in eq.~\eqref{eq:sde1b} follows to 
\begin{align}
B_{ij}=O_{im}\Lambda^{1/2}_{mk}O^T_{kj}\,. 
\end{align}
Thus, the diagonal and off-diagonal elements $B_{ij}$ in eqns.~\eqref{eq:sde2a} and \eqref{eq:sde2c} can be determined. In the following, we summarize the relevant stochastic model coefficients for this classical framework, following ref.~\cite{Pope2002}. The model equations are integrated by a standard Euler-Murayama scheme.   

\section{Flow characteristics}
\label{sec:Flow characteristics}
The underlying HSF for this investigation is described in ref.~\cite{Pope2002}. In the following, we compactly summarize the considered flow characteristics. For the characteristic scales see Table~\ref{tab:Characteristic scales}. All the following flow parameters are given in dimensionless way. Dimensionless Reynolds stress tensor, diffusion matrix and covariance matrix are given by
\begin{align}
    \tilde{C} &= \left[\begin{array}{ccc}
        4.43  &  0    & -1.48\\
        0     & 2.81  & 0    \\
        -1.48 & 0     & 1.98 \\
    \end{array}\right], \\
    \tilde{B} &= \left[\begin{array}{ccc}
        3.03  &  0    & -0.47\\
        0     & 3.50  & 0    \\
        -0.47 & 0     & 2.98 \\
    \end{array}\right], \\
    \tilde{\Sigma} &= \tilde{B}\tilde{B}^Td\tilde{t} = \left[\begin{array}{ccc}
         0.09  &  0    & -0.03\\
         0     & 0.12  & 0    \\
        -0.03  & 0     & 0.09 \\
    \end{array}\right],
\end{align}
respectively. The dimensionless Lagrangian integral time scale is denoted as
\begin{align}
    \tilde{\tau} &= \left(\begin{array}{ccc}
        0.84,  &  0.46,    & 0.42\\
    \end{array}\right)^T.
\end{align}

\begin{table}[]
    \centering    
    \begin{tabular}{lcr}
        \hline\hline
        Parameter & Variable & Value \\
        \hline
        Shear rate &$S$& $3.51$  \\
        Mean dissipation rate& $\langle \varepsilon \rangle$ & $1.34$ \\
        Turbulent kinetic energy & $k$ & $1.84$ \\
        Characteristic time & $T_S$ & $0.28$ \\
        Characteristic length & $L_S$ & $0.18$ \\
        Characteristic velocity & $U_S$ & $0.62$ \\
        \hline\hline
    \end{tabular}
    \caption{Characteristic scales of the homogeneous shear flow case. Data are taken from ref. \cite{Pope2002}.}
    \label{tab:Characteristic scales}
\end{table}

\section{Classical Markov-chain Monte Carlo sampling}
\label{sec:MCMC}
Markov-chain Monte Carlo (MCMC) methods are a class of algorithms used for generating samples from complex, high-dimensional probability distributions. A compact introduction can be found in the textbook by Bishop.\cite{Bishop2006} They construct a Markov chain whose stationary distribution corresponds to a given target distribution $p(\bm u^\prime)$. By iteratively evolving the chain, one obtains a sequence of correlated samples that asymptotically follow $p$. This allows for the approximation of expectations and statistical properties of $p$, even when direct sampling is not feasible.

In this work, the Metropolis-Hastings algorithm is employed to generate samples of turbulent velocity fluctuations from a known target distribution. Starting from a current state $\bm u^{\prime\,(m)}$, a new proposal $\bm u^*$ is drawn from a symmetric Gaussian proposal distribution $q(\bm u^*\mid\bm u^{\prime\,(m)})\sim\mathcal{N}(\bm u^{\prime\,(m)}, \sigma_\mathrm{prop}^2\mathbf I)$, where $\sigma_\mathrm{prop}^2$ is the proposal variance and $\mathbf{I}$ denotes the identity matrix in $\mathbb{R}^3$. The proposal is then accepted with probability
\begin{align}
    \alpha(\bm u^*\mid\bm u^{\prime\,(m)}) = \min\left\{1, \frac{p(\bm u^*)}{p(\bm u^{\prime\,(m)})}\right\},
\end{align}
exploiting the symmetry $q(\bm u^*\mid\bm u^{\prime\,(m)}) = q(\bm u^{\prime\,(m)}\mid\bm u^*)$. Otherwise, the previous state is retained. This accept–reject rule preserves the detailed balance condition, ensuring that the Markov chain converges to the correct equilibrium distribution, and therefore enables accurate sampling from the target distribution $p$.

To select an optimal value for $\sigma_\mathrm{prop}$, we minimized the mean squared logarithmic error (MSLE) between the dispersion curve $D_\mathrm{MCMC}(t)$ obtained from the MCMC simulation and the reference curve $D_\mathrm{SDE}(t)$ computed from the stochastic differential equation model,
\begin{align}
    \mathrm{MSLE} = \frac{1}{M_t}\sum_{m=0}^{M_t-1}\left[\log\!\left(1 + D_\mathrm{SDE}(\tilde t_m)\right) - \log\!\left(1 + D_\mathrm{MCMC}(\tilde t_m)\right)\right]^2\!,
\end{align}
where $\tilde t_m = m\,d\tilde t$, and $M_t$ denotes the number of time steps used in the simulation. A polynomial fit of the resulting MSLE curve yielded a minimum at $\sigma_\mathrm{prop}^* = 0.400$, corresponding to a minimal MSLE of $0.142$ with a coefficient of determination of $R^2 = 0.998$. This value was subsequently used in all MCMC simulations.

\section{One-dimensional random walk with IQM Garnet}
\label{sec:1dRandomWalkIQM}
In this section, we analyze tracer transport using the proposed hybrid classical–quantum method, where particle paths are computed using noise terms from a real quantum device. For simplicity, we consider a one--dimensional random walk mentioned in ref.~\cite{Goetzfried2019}. The Euler-Maruyama scheme is used to compute the particle positions $x(t)$ such that $x(t+dt) = x(t) + \xi$, where the noise term $\xi$ is drawn from a normal distribution $p(\xi)$ with variance $\sigma^2 = 2D dt$. Here, $D$ is a prescribed diffusion constant. In the classical case, this corresponds to $\xi = \sqrt{2D}dW$. 

\begin{figure}
\centering
\includegraphics[width=0.45\textwidth]{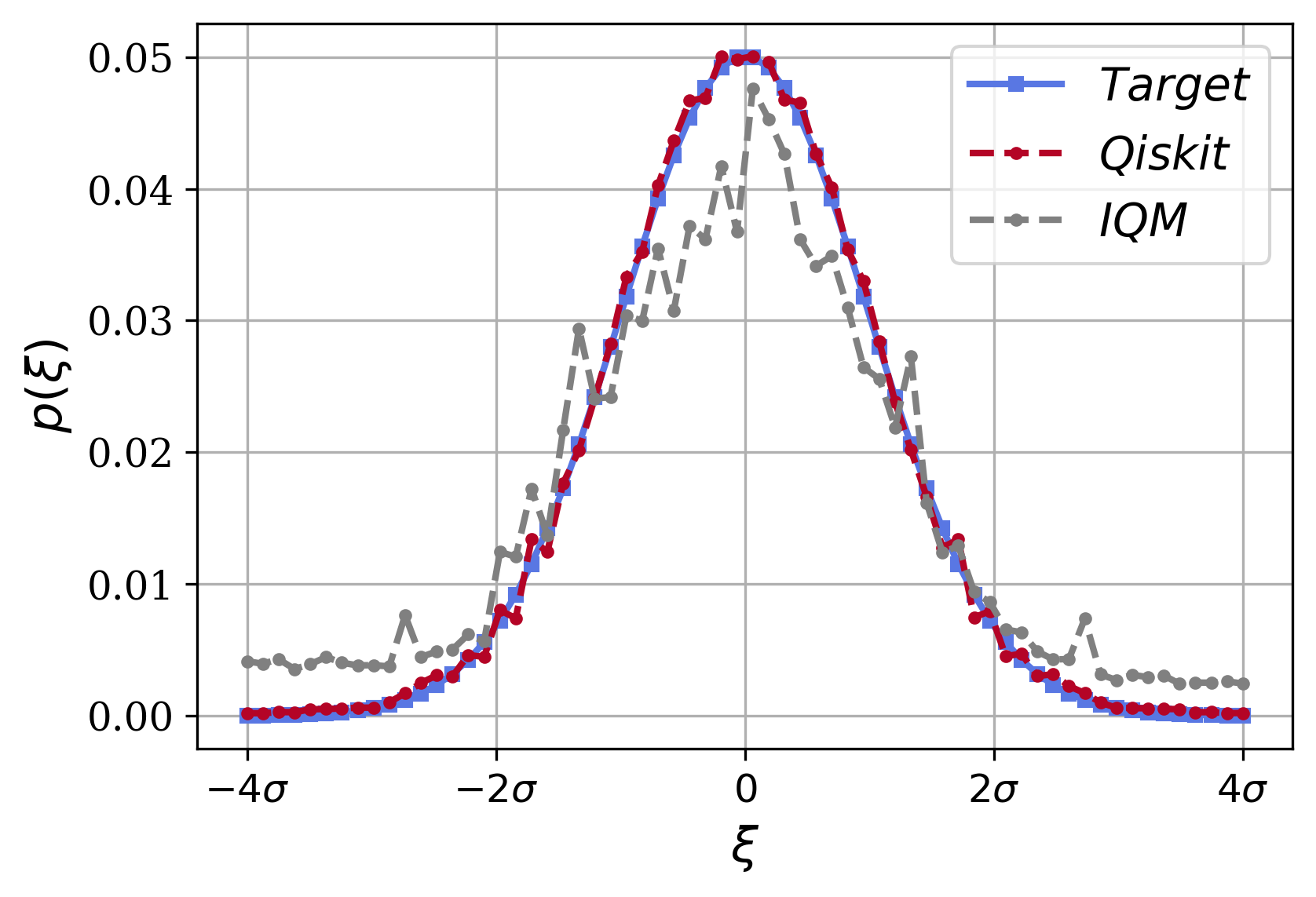}   
\caption{Reconstructed PDFs for the mean evaluation of $100$ times $20000$ shots with $n=6$ qubits, where the result of the Qiskit Statevector Sampler (red) and the result of the real IQM Garnet (grey) are compared to the target PDF (blue).}
\label{fig:1dRandomWalk_Reconstruction}
\end{figure}

The PDF $p(\xi)$ is reconstructed using a $n=6$ qubit DVG with a $d=3$ layer linear circuit architecture. Figure~\ref{fig:1dRandomWalk_Reconstruction} compares the reconstruction using the IQM Garnet QPU to results from the Qiskit Statevector Sampler and the target distribution.  The IQM Garnet struggles to accurately reproduce both the peak and the tails of the distribution, unlike the ideal statevector simulator. 

We obtain one-shot measurements by a batched execution on real hardware to minimize overhead from queuing and communication on real hardware. We execute $20000$ shots per job (maximum) and post-select a measurement per particle path, leading to as many queued executions as time steps.

\begin{figure}
\centering
\includegraphics[width=0.45\textwidth]{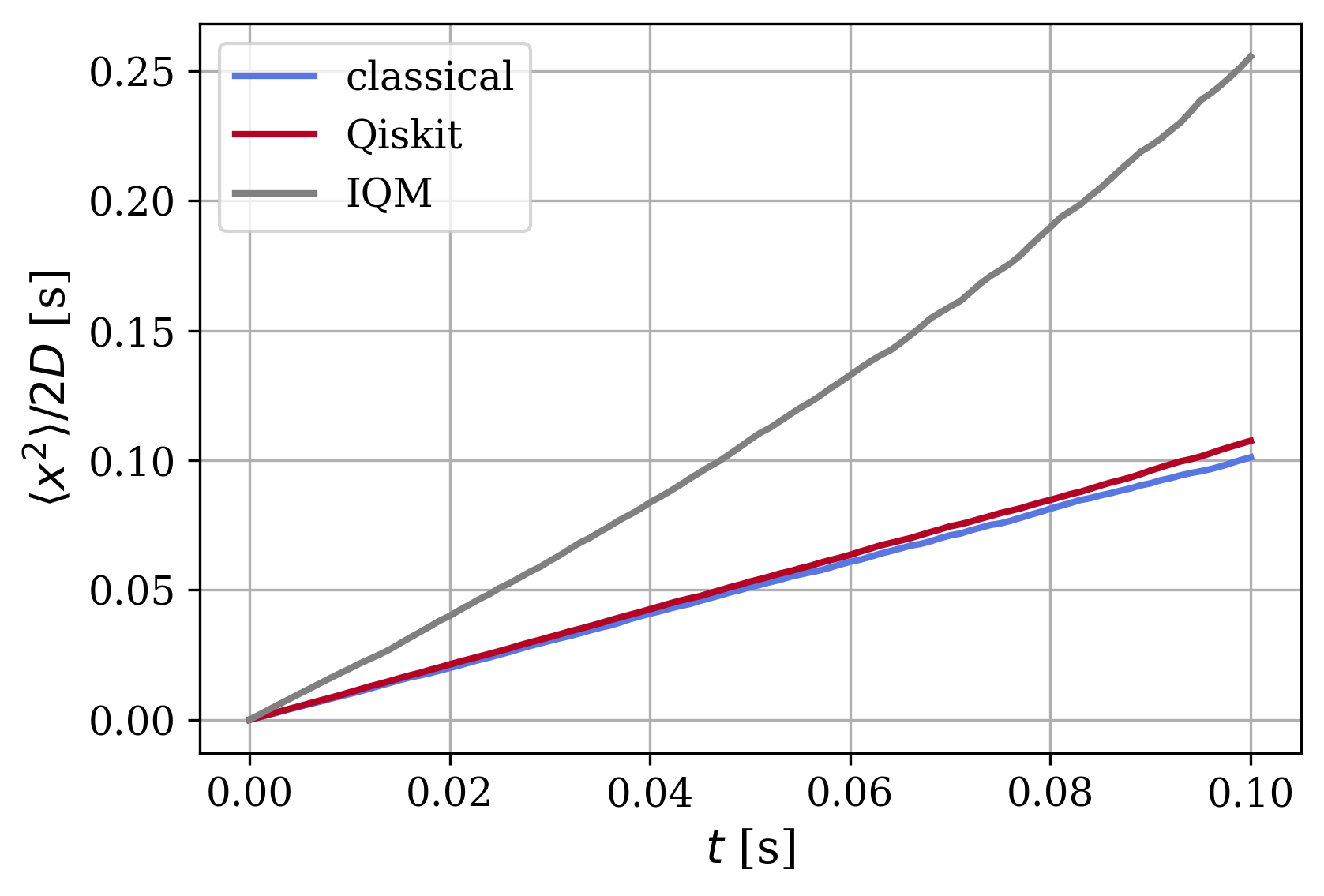}   
\caption{Temporal evolution of the mean squared displacement for a one-dimensional Brownian walk. We compare results of the Qiskit Statevector Sampler (red), the real IQM Garnet (grey) with the classical computation (blue). Data are obtained as an average over 20000 trajectories in each case.}
\label{fig:1dRandomWalk_x2}
\end{figure}

For the analysis, we evaluate the mean squared displacement $\langle x ^2 \rangle$ over time $t$ as shown in Fig.~\ref{fig:1dRandomWalk_x2}. The results from the Qiskit Statevector Sampler align well with the classical curve. In contrast, the IQM Garnet shows significant deviation and a nonlinear growth. This discrepancy results from the gate and readout errors at the real quantum device.


\bibliography{DVG}

\begin{thebibliography}{67}%
\makeatletter
\providecommand \@ifxundefined [1]{%
 \@ifx{#1\undefined}
}%
\providecommand \@ifnum [1]{%
 \ifnum #1\expandafter \@firstoftwo
 \else \expandafter \@secondoftwo
 \fi
}%
\providecommand \@ifx [1]{%
 \ifx #1\expandafter \@firstoftwo
 \else \expandafter \@secondoftwo
 \fi
}%
\providecommand \natexlab [1]{#1}%
\providecommand \enquote  [1]{``#1''}%
\providecommand \bibnamefont  [1]{#1}%
\providecommand \bibfnamefont [1]{#1}%
\providecommand \citenamefont [1]{#1}%
\providecommand \href@noop [0]{\@secondoftwo}%
\providecommand \href [0]{\begingroup \@sanitize@url \@href}%
\providecommand \@href[1]{\@@startlink{#1}\@@href}%
\providecommand \@@href[1]{\endgroup#1\@@endlink}%
\providecommand \@sanitize@url [0]{\catcode `\\12\catcode `\$12\catcode
  `\&12\catcode `\#12\catcode `\^12\catcode `\_12\catcode `\%12\relax}%
\providecommand \@@startlink[1]{}%
\providecommand \@@endlink[0]{}%
\providecommand \url  [0]{\begingroup\@sanitize@url \@url }%
\providecommand \@url [1]{\endgroup\@href {#1}{\urlprefix }}%
\providecommand \urlprefix  [0]{URL }%
\providecommand \Eprint [0]{\href }%
\providecommand \doibase [0]{http://dx.doi.org/}%
\providecommand \selectlanguage [0]{\@gobble}%
\providecommand \bibinfo  [0]{\@secondoftwo}%
\providecommand \bibfield  [0]{\@secondoftwo}%
\providecommand \translation [1]{[#1]}%
\providecommand \BibitemOpen [0]{}%
\providecommand \bibitemStop [0]{}%
\providecommand \bibitemNoStop [0]{.\EOS\space}%
\providecommand \EOS [0]{\spacefactor3000\relax}%
\providecommand \BibitemShut  [1]{\csname bibitem#1\endcsname}%
\let\auto@bib@innerbib\@empty
\bibitem [{\citenamefont {Preskill}(2018)}]{Preskill2018}%
  \BibitemOpen
  \bibfield  {author} {\bibinfo {author} {\bibfnamefont {J.}~\bibnamefont
  {Preskill}},\ }\bibfield  {title} {\enquote {\bibinfo {title} {{Quantum
  computing in the NISQ era and beyond}},}\ }\href@noop {} {\bibfield
  {journal} {\bibinfo  {journal} {Quantum}\ }\textbf {\bibinfo {volume} {2}},\
  \bibinfo {pages} {79} (\bibinfo {year} {2018})}\BibitemShut {NoStop}%
\bibitem [{\citenamefont {Deutsch}(2020)}]{Deutsch2020}%
  \BibitemOpen
  \bibfield  {author} {\bibinfo {author} {\bibfnamefont {I.~H.}\ \bibnamefont
  {Deutsch}},\ }\bibfield  {title} {\enquote {\bibinfo {title} {{Harnessing the
  power of the second quantum revolution}},}\ }\href@noop {} {\bibfield
  {journal} {\bibinfo  {journal} {PRX Quantum}\ }\textbf {\bibinfo {volume}
  {1}},\ \bibinfo {pages} {020101} (\bibinfo {year} {2020})}\BibitemShut
  {NoStop}%
\bibitem [{\citenamefont {Bharadwaj}\ and\ \citenamefont
  {Sreenivasan}(2020)}]{Bharadwaj2020}%
  \BibitemOpen
  \bibfield  {author} {\bibinfo {author} {\bibfnamefont {S.~S.}\ \bibnamefont
  {Bharadwaj}}\ and\ \bibinfo {author} {\bibfnamefont {K.~R.}\ \bibnamefont
  {Sreenivasan}},\ }\bibfield  {title} {\enquote {\bibinfo {title} {Quantum
  computation of fluid dynamics},}\ }\href@noop {} {\bibfield  {journal}
  {\bibinfo  {journal} {Indian Academy of Sciences Conference Series}\ }\textbf
  {\bibinfo {volume} {3}},\ \bibinfo {pages} {77--96} (\bibinfo {year}
  {2020})}\BibitemShut {NoStop}%
\bibitem [{\citenamefont {Lubasch}\ \emph {et~al.}(2020)\citenamefont
  {Lubasch}, \citenamefont {Joo}, \citenamefont {Moinier}, \citenamefont
  {Kiffner},\ and\ \citenamefont {Jaksch}}]{Lubasch2020}%
  \BibitemOpen
  \bibfield  {author} {\bibinfo {author} {\bibfnamefont {M.}~\bibnamefont
  {Lubasch}}, \bibinfo {author} {\bibfnamefont {J.}~\bibnamefont {Joo}},
  \bibinfo {author} {\bibfnamefont {P.}~\bibnamefont {Moinier}}, \bibinfo
  {author} {\bibfnamefont {M.}~\bibnamefont {Kiffner}}, \ and\ \bibinfo
  {author} {\bibfnamefont {D.}~\bibnamefont {Jaksch}},\ }\bibfield  {title}
  {\enquote {\bibinfo {title} {Variational quantum algorithms for nonlinear
  problems},}\ }\href {\doibase 10.1103/PhysRevA.101.010301} {\bibfield
  {journal} {\bibinfo  {journal} {Phys. Rev. A}\ }\textbf {\bibinfo {volume}
  {101}},\ \bibinfo {pages} {010301} (\bibinfo {year} {2020})}\BibitemShut
  {NoStop}%
\bibitem [{\citenamefont {Succi}\ \emph {et~al.}(2023)\citenamefont {Succi},
  \citenamefont {Itani}, \citenamefont {Sreenivasan},\ and\ \citenamefont
  {Steijl}}]{Succi2023a}%
  \BibitemOpen
  \bibfield  {author} {\bibinfo {author} {\bibfnamefont {S.}~\bibnamefont
  {Succi}}, \bibinfo {author} {\bibfnamefont {W.}~\bibnamefont {Itani}},
  \bibinfo {author} {\bibfnamefont {K.~R.}\ \bibnamefont {Sreenivasan}}, \ and\
  \bibinfo {author} {\bibfnamefont {R.}~\bibnamefont {Steijl}},\ }\bibfield
  {title} {\enquote {\bibinfo {title} {Quantum computing for fluids: Where do
  we stand?}}\ }\href@noop {} {\bibfield  {journal} {\bibinfo  {journal}
  {Europhys. Lett.}\ }\textbf {\bibinfo {volume} {144}},\ \bibinfo {pages}
  {{10001}} (\bibinfo {year} {2023})}\BibitemShut {NoStop}%
\bibitem [{\citenamefont {Ingelmann}\ \emph {et~al.}(2024)\citenamefont
  {Ingelmann}, \citenamefont {Bharadwaj}, \citenamefont {Pfeffer},
  \citenamefont {Sreenivasan},\ and\ \citenamefont
  {Schumacher}}]{Ingelmann2024}%
  \BibitemOpen
  \bibfield  {author} {\bibinfo {author} {\bibfnamefont {J.}~\bibnamefont
  {Ingelmann}}, \bibinfo {author} {\bibfnamefont {S.~S.}\ \bibnamefont
  {Bharadwaj}}, \bibinfo {author} {\bibfnamefont {P.}~\bibnamefont {Pfeffer}},
  \bibinfo {author} {\bibfnamefont {K.~R.}\ \bibnamefont {Sreenivasan}}, \ and\
  \bibinfo {author} {\bibfnamefont {J.}~\bibnamefont {Schumacher}},\ }\bibfield
   {title} {\enquote {\bibinfo {title} {{Two quantum algorithms for solving the
  one-dimensional advection–diffusion equation}},}\ }\href@noop {} {\bibfield
   {journal} {\bibinfo  {journal} {Comput. Fluids}\ }\textbf {\bibinfo {volume}
  {281}},\ \bibinfo {pages} {106369} (\bibinfo {year} {2024})}\BibitemShut
  {NoStop}%
\bibitem [{\citenamefont {Tennie}\ \emph {et~al.}(2025)\citenamefont {Tennie},
  \citenamefont {Laizet}, \citenamefont {Lloyd},\ and\ \citenamefont
  {Magri}}]{Tennie2025}%
  \BibitemOpen
  \bibfield  {author} {\bibinfo {author} {\bibfnamefont {F.}~\bibnamefont
  {Tennie}}, \bibinfo {author} {\bibfnamefont {S.}~\bibnamefont {Laizet}},
  \bibinfo {author} {\bibfnamefont {S.}~\bibnamefont {Lloyd}}, \ and\ \bibinfo
  {author} {\bibfnamefont {L.}~\bibnamefont {Magri}},\ }\bibfield  {title}
  {\enquote {\bibinfo {title} {Quantum computing for nonlinear differential
  equations and turbulence},}\ }\href {\doibase 10.1038/s42254-024-00799-w}
  {\bibfield  {journal} {\bibinfo  {journal} {Nat. Rev. Phys.}\ }\textbf
  {\bibinfo {volume} {7}},\ \bibinfo {pages} {1--11} (\bibinfo {year}
  {2025})}\BibitemShut {NoStop}%
\bibitem [{\citenamefont {Pfeffer}\ \emph {et~al.}(2025)\citenamefont
  {Pfeffer}, \citenamefont {Brearley}, \citenamefont {Laizet},\ and\
  \citenamefont {Schumacher}}]{Pfeffer2025}%
  \BibitemOpen
  \bibfield  {author} {\bibinfo {author} {\bibfnamefont {P.}~\bibnamefont
  {Pfeffer}}, \bibinfo {author} {\bibfnamefont {P.}~\bibnamefont {Brearley}},
  \bibinfo {author} {\bibfnamefont {S.}~\bibnamefont {Laizet}}, \ and\ \bibinfo
  {author} {\bibfnamefont {J.}~\bibnamefont {Schumacher}},\ }\href
  {https://arxiv.org/abs/2505.10136} {\enquote {\bibinfo {title} {Spectral
  quantum algorithm for passive scalar transport in shear flows},}\ } (\bibinfo
  {year} {2025}),\ \Eprint {http://arxiv.org/abs/2505.10136} {arXiv:2505.10136
  [quant-ph]} \BibitemShut {NoStop}%
\bibitem [{\citenamefont {Biamonte}\ \emph {et~al.}(2017)\citenamefont
  {Biamonte}, \citenamefont {Wittek}, \citenamefont {Pancotti}, \citenamefont
  {Wiebe},\ and\ \citenamefont {Lloyd}}]{Biamonte2017}%
  \BibitemOpen
  \bibfield  {author} {\bibinfo {author} {\bibfnamefont {J.}~\bibnamefont
  {Biamonte}}, \bibinfo {author} {\bibfnamefont {P.}~\bibnamefont {Wittek}},
  \bibinfo {author} {\bibfnamefont {N.}~\bibnamefont {Pancotti}}, \bibinfo
  {author} {\bibfnamefont {N.}~\bibnamefont {Wiebe}}, \ and\ \bibinfo {author}
  {\bibfnamefont {S.}~\bibnamefont {Lloyd}},\ }\bibfield  {title} {\enquote
  {\bibinfo {title} {Quantum machine learning},}\ }\href@noop {} {\bibfield
  {journal} {\bibinfo  {journal} {Nature}\ }\textbf {\bibinfo {volume} {549}},\
  \bibinfo {pages} {195--202} (\bibinfo {year} {2017})}\BibitemShut {NoStop}%
\bibitem [{\citenamefont {Fujii}\ and\ \citenamefont
  {Nakajima}(2017)}]{Fujii2017}%
  \BibitemOpen
  \bibfield  {author} {\bibinfo {author} {\bibfnamefont {K.}~\bibnamefont
  {Fujii}}\ and\ \bibinfo {author} {\bibfnamefont {K.}~\bibnamefont
  {Nakajima}},\ }\bibfield  {title} {\enquote {\bibinfo {title} {Harnessing
  disordered-ensemble quantum dynamics for machine learning},}\ }\href
  {\doibase 10.1103/PhysRevApplied.8.024030} {\bibfield  {journal} {\bibinfo
  {journal} {Phys. Rev. Applied}\ }\textbf {\bibinfo {volume} {8}},\ \bibinfo
  {pages} {024030} (\bibinfo {year} {2017})}\BibitemShut {NoStop}%
\bibitem [{\citenamefont {Chen}, \citenamefont {Nurdin},\ and\ \citenamefont
  {Yamamoto}(2020)}]{Chen2020}%
  \BibitemOpen
  \bibfield  {author} {\bibinfo {author} {\bibfnamefont {J.}~\bibnamefont
  {Chen}}, \bibinfo {author} {\bibfnamefont {H.~I.}\ \bibnamefont {Nurdin}}, \
  and\ \bibinfo {author} {\bibfnamefont {N.}~\bibnamefont {Yamamoto}},\
  }\bibfield  {title} {\enquote {\bibinfo {title} {Temporal information
  processing on noisy quantum computers},}\ }\href {\doibase
  10.1103/PhysRevApplied.14.024065} {\bibfield  {journal} {\bibinfo  {journal}
  {Phys. Rev. Applied}\ }\textbf {\bibinfo {volume} {14}},\ \bibinfo {pages}
  {024065} (\bibinfo {year} {2020})}\BibitemShut {NoStop}%
\bibitem [{\citenamefont {Pfeffer}, \citenamefont {Heyder},\ and\ \citenamefont
  {Schumacher}(2023)}]{Pfeffer2023}%
  \BibitemOpen
  \bibfield  {author} {\bibinfo {author} {\bibfnamefont {P.}~\bibnamefont
  {Pfeffer}}, \bibinfo {author} {\bibfnamefont {F.}~\bibnamefont {Heyder}}, \
  and\ \bibinfo {author} {\bibfnamefont {J.}~\bibnamefont {Schumacher}},\
  }\bibfield  {title} {\enquote {\bibinfo {title} {Reduced-order modeling of
  two-dimensional turbulent {R}ayleigh-{B}{\'{e}}nard flow by hybrid
  quantum-classical reservoir computing},}\ }\href@noop {} {\bibfield
  {journal} {\bibinfo  {journal} {Phys. Rev. Research}\ }\textbf {\bibinfo
  {volume} {5}},\ \bibinfo {pages} {043242} (\bibinfo {year}
  {2023})}\BibitemShut {NoStop}%
\bibitem [{\citenamefont {Ahmed}, \citenamefont {Tennie},\ and\ \citenamefont
  {Magri}(2024)}]{Ahmed2024}%
  \BibitemOpen
  \bibfield  {author} {\bibinfo {author} {\bibfnamefont {O.}~\bibnamefont
  {Ahmed}}, \bibinfo {author} {\bibfnamefont {F.}~\bibnamefont {Tennie}}, \
  and\ \bibinfo {author} {\bibfnamefont {L.}~\bibnamefont {Magri}},\ }\bibfield
   {title} {\enquote {\bibinfo {title} {Prediction of chaotic dynamics and
  extreme events: A recurrence-free quantum reservoir computing approach},}\
  }\href {\doibase 10.1103/PhysRevResearch.6.043082} {\bibfield  {journal}
  {\bibinfo  {journal} {Phys. Rev. Res.}\ }\textbf {\bibinfo {volume} {6}},\
  \bibinfo {pages} {043082} (\bibinfo {year} {2024})}\BibitemShut {NoStop}%
\bibitem [{\citenamefont {Yeung}(2002)}]{Yeung2002}%
  \BibitemOpen
  \bibfield  {author} {\bibinfo {author} {\bibfnamefont {P.~K.}\ \bibnamefont
  {Yeung}},\ }\bibfield  {title} {\enquote {\bibinfo {title} {{Lagrangian
  investigations of turbulence}},}\ }\href@noop {} {\bibfield  {journal}
  {\bibinfo  {journal} {Annu. Rev. Fluid Mech.}\ }\textbf {\bibinfo {volume}
  {34}},\ \bibinfo {pages} {115--142} (\bibinfo {year} {2002})}\BibitemShut
  {NoStop}%
\bibitem [{\citenamefont {Toschi}\ and\ \citenamefont
  {Bodenschatz}(2009)}]{Toschi2009}%
  \BibitemOpen
  \bibfield  {author} {\bibinfo {author} {\bibfnamefont {F.}~\bibnamefont
  {Toschi}}\ and\ \bibinfo {author} {\bibfnamefont {E.}~\bibnamefont
  {Bodenschatz}},\ }\bibfield  {title} {\enquote {\bibinfo {title} {{Lagrangian
  properties of particles in turbulence}},}\ }\href@noop {} {\bibfield
  {journal} {\bibinfo  {journal} {Annu. Rev. Fluid Mech.}\ }\textbf {\bibinfo
  {volume} {41}},\ \bibinfo {pages} {375--404} (\bibinfo {year}
  {2009})}\BibitemShut {NoStop}%
\bibitem [{\citenamefont {Worm}\ \emph {et~al.}(2017)\citenamefont {Worm},
  \citenamefont {Lotze}, \citenamefont {Jubinville}, \citenamefont {Wilcox},\
  and\ \citenamefont {Jambeck}}]{Worm2017}%
  \BibitemOpen
  \bibfield  {author} {\bibinfo {author} {\bibfnamefont {B.}~\bibnamefont
  {Worm}}, \bibinfo {author} {\bibfnamefont {H.~K.}\ \bibnamefont {Lotze}},
  \bibinfo {author} {\bibfnamefont {I.}~\bibnamefont {Jubinville}}, \bibinfo
  {author} {\bibfnamefont {C.}~\bibnamefont {Wilcox}}, \ and\ \bibinfo {author}
  {\bibfnamefont {J.}~\bibnamefont {Jambeck}},\ }\bibfield  {title} {\enquote
  {\bibinfo {title} {{Plastic as a persistent marine pollutant}},}\ }\href@noop
  {} {\bibfield  {journal} {\bibinfo  {journal} {Annu. Rev. Environ. Resour.}\
  }\textbf {\bibinfo {volume} {42}},\ \bibinfo {pages} {1--26} (\bibinfo {year}
  {2017})}\BibitemShut {NoStop}%
\bibitem [{\citenamefont {Zhao}\ \emph {et~al.}(2025)\citenamefont {Zhao},
  \citenamefont {Kvale}, \citenamefont {Zhu}, \citenamefont {Zettler},
  \citenamefont {Egger}, \citenamefont {Mincer}, \citenamefont
  {Amaral-Zettler}, \citenamefont {Lebreton}, \citenamefont {Niemann},
  \citenamefont {Nakajima}, \citenamefont {Thiel}, \citenamefont {Bos},
  \citenamefont {Galgani},\ and\ \citenamefont {Stubbins}}]{Zhao2025}%
  \BibitemOpen
  \bibfield  {author} {\bibinfo {author} {\bibfnamefont {S.}~\bibnamefont
  {Zhao}}, \bibinfo {author} {\bibfnamefont {K.~F.}\ \bibnamefont {Kvale}},
  \bibinfo {author} {\bibfnamefont {L.}~\bibnamefont {Zhu}}, \bibinfo {author}
  {\bibfnamefont {E.~R.}\ \bibnamefont {Zettler}}, \bibinfo {author}
  {\bibfnamefont {M.}~\bibnamefont {Egger}}, \bibinfo {author} {\bibfnamefont
  {T.~J.}\ \bibnamefont {Mincer}}, \bibinfo {author} {\bibfnamefont {L.~A.}\
  \bibnamefont {Amaral-Zettler}}, \bibinfo {author} {\bibfnamefont
  {L.}~\bibnamefont {Lebreton}}, \bibinfo {author} {\bibfnamefont
  {H.}~\bibnamefont {Niemann}}, \bibinfo {author} {\bibfnamefont
  {R.}~\bibnamefont {Nakajima}}, \bibinfo {author} {\bibfnamefont
  {M.}~\bibnamefont {Thiel}}, \bibinfo {author} {\bibfnamefont {R.~P.}\
  \bibnamefont {Bos}}, \bibinfo {author} {\bibfnamefont {L.}~\bibnamefont
  {Galgani}}, \ and\ \bibinfo {author} {\bibfnamefont {A.}~\bibnamefont
  {Stubbins}},\ }\bibfield  {title} {\enquote {\bibinfo {title} {The
  distribution of subsurface microplastics in the ocean},}\ }\href {\doibase
  10.1038/s41586-025-08818-1} {\bibfield  {journal} {\bibinfo  {journal}
  {Nature}\ }\textbf {\bibinfo {volume} {641}},\ \bibinfo {pages} {51--61}
  (\bibinfo {year} {2025})}\BibitemShut {NoStop}%
\bibitem [{\citenamefont {Ram}\ and\ \citenamefont {Gayley}(1991)}]{Ram1991}%
  \BibitemOpen
  \bibfield  {author} {\bibinfo {author} {\bibfnamefont {M.}~\bibnamefont
  {Ram}}\ and\ \bibinfo {author} {\bibfnamefont {R.~I.}\ \bibnamefont
  {Gayley}},\ }\bibfield  {title} {\enquote {\bibinfo {title} {{Long-range
  transport of volcanic ash to the Greenland ice sheet}},}\ }\href@noop {}
  {\bibfield  {journal} {\bibinfo  {journal} {Nature}\ }\textbf {\bibinfo
  {volume} {349}},\ \bibinfo {pages} {401--404} (\bibinfo {year}
  {1991})}\BibitemShut {NoStop}%
\bibitem [{\citenamefont {Woods}(2010)}]{Woods2010}%
  \BibitemOpen
  \bibfield  {author} {\bibinfo {author} {\bibfnamefont {A.~W.}\ \bibnamefont
  {Woods}},\ }\bibfield  {title} {\enquote {\bibinfo {title} {{Turbulent plumes
  in nature}},}\ }\href@noop {} {\bibfield  {journal} {\bibinfo  {journal}
  {Annu. Rev. Fluid Mech.}\ }\textbf {\bibinfo {volume} {42}},\ \bibinfo
  {pages} {391--412} (\bibinfo {year} {2010})}\BibitemShut {NoStop}%
\bibitem [{\citenamefont {Arnalds}\ \emph {et~al.}(2013)\citenamefont
  {Arnalds}, \citenamefont {Thorarinsdottir}, \citenamefont {Thorsson},
  \citenamefont {Waldhauserova},\ and\ \citenamefont
  {Agustsdottir}}]{Arnalds2013}%
  \BibitemOpen
  \bibfield  {author} {\bibinfo {author} {\bibfnamefont {O.}~\bibnamefont
  {Arnalds}}, \bibinfo {author} {\bibfnamefont {E.~F.}\ \bibnamefont
  {Thorarinsdottir}}, \bibinfo {author} {\bibfnamefont {J.}~\bibnamefont
  {Thorsson}}, \bibinfo {author} {\bibfnamefont {P.~D.}\ \bibnamefont
  {Waldhauserova}}, \ and\ \bibinfo {author} {\bibfnamefont {A.~M.}\
  \bibnamefont {Agustsdottir}},\ }\bibfield  {title} {\enquote {\bibinfo
  {title} {{An extreme wind erosion event of the fresh Eyjafjallajökull 2010
  volcanic ash}},}\ }\href@noop {} {\bibfield  {journal} {\bibinfo  {journal}
  {Sci. Rep.}\ }\textbf {\bibinfo {volume} {3}},\ \bibinfo {pages} {1257}
  (\bibinfo {year} {2013})}\BibitemShut {NoStop}%
\bibitem [{\citenamefont {Hamaoui-Laguel}\ \emph {et~al.}(2015)\citenamefont
  {Hamaoui-Laguel}, \citenamefont {Vautard}, \citenamefont {Liu}, \citenamefont
  {Solmon}, \citenamefont {Viovy}, \citenamefont {Khvorostyanov}, \citenamefont
  {Essl}, \citenamefont {Chuine}, \citenamefont {Colette}, \citenamefont
  {Semenov}, \citenamefont {Schaffhauser}, \citenamefont {Storkey},
  \citenamefont {Thibaudon},\ and\ \citenamefont {Epstein}}]{Hamaoui2015}%
  \BibitemOpen
  \bibfield  {author} {\bibinfo {author} {\bibfnamefont {L.}~\bibnamefont
  {Hamaoui-Laguel}}, \bibinfo {author} {\bibfnamefont {R.}~\bibnamefont
  {Vautard}}, \bibinfo {author} {\bibfnamefont {L.}~\bibnamefont {Liu}},
  \bibinfo {author} {\bibfnamefont {F.}~\bibnamefont {Solmon}}, \bibinfo
  {author} {\bibfnamefont {N.}~\bibnamefont {Viovy}}, \bibinfo {author}
  {\bibfnamefont {D.}~\bibnamefont {Khvorostyanov}}, \bibinfo {author}
  {\bibfnamefont {F.}~\bibnamefont {Essl}}, \bibinfo {author} {\bibfnamefont
  {I.}~\bibnamefont {Chuine}}, \bibinfo {author} {\bibfnamefont
  {A.}~\bibnamefont {Colette}}, \bibinfo {author} {\bibfnamefont {M.~A.}\
  \bibnamefont {Semenov}}, \bibinfo {author} {\bibfnamefont {A.}~\bibnamefont
  {Schaffhauser}}, \bibinfo {author} {\bibfnamefont {J.}~\bibnamefont
  {Storkey}}, \bibinfo {author} {\bibfnamefont {M.}~\bibnamefont {Thibaudon}},
  \ and\ \bibinfo {author} {\bibfnamefont {M.~M.}\ \bibnamefont {Epstein}},\
  }\bibfield  {title} {\enquote {\bibinfo {title} {{Effects of climate change
  and seed dispersal on airborne ragweed pollen loads in Europe}},}\
  }\href@noop {} {\bibfield  {journal} {\bibinfo  {journal} {Nat. Climate
  Change}\ }\textbf {\bibinfo {volume} {5}},\ \bibinfo {pages} {766--771}
  (\bibinfo {year} {2015})}\BibitemShut {NoStop}%
\bibitem [{\citenamefont {Meneveau}(2011)}]{Meneveau2011}%
  \BibitemOpen
  \bibfield  {author} {\bibinfo {author} {\bibfnamefont {C.}~\bibnamefont
  {Meneveau}},\ }\bibfield  {title} {\enquote {\bibinfo {title} {Lagrangian
  dynamics and models of the velocity gradient tensor in turbulent flows},}\
  }\href@noop {} {\bibfield  {journal} {\bibinfo  {journal} {Annu. Rev. Fluid
  Mech.}\ }\textbf {\bibinfo {volume} {43}},\ \bibinfo {pages} {219--245}
  (\bibinfo {year} {2011})}\BibitemShut {NoStop}%
\bibitem [{\citenamefont {Götzfried}\ \emph {et~al.}(2019)\citenamefont
  {Götzfried}, \citenamefont {Emran}, \citenamefont {Villermaux},\ and\
  \citenamefont {Schumacher}}]{Goetzfried2019}%
  \BibitemOpen
  \bibfield  {author} {\bibinfo {author} {\bibfnamefont {P.}~\bibnamefont
  {Götzfried}}, \bibinfo {author} {\bibfnamefont {M.~S.}\ \bibnamefont
  {Emran}}, \bibinfo {author} {\bibfnamefont {E.}~\bibnamefont {Villermaux}}, \
  and\ \bibinfo {author} {\bibfnamefont {J.}~\bibnamefont {Schumacher}},\
  }\bibfield  {title} {\enquote {\bibinfo {title} {Comparison of lagrangian and
  eulerian frames of passive scalar turbulent mixing},}\ }\href@noop {}
  {\bibfield  {journal} {\bibinfo  {journal} {Phys. Rev. Fluids}\ }\textbf
  {\bibinfo {volume} {4}},\ \bibinfo {pages} {044607} (\bibinfo {year}
  {2019})}\BibitemShut {NoStop}%
\bibitem [{\citenamefont {Lin}\ \emph {et~al.}(2003)\citenamefont {Lin},
  \citenamefont {Gerbig}, \citenamefont {Wofsy}, \citenamefont {Andrews},
  \citenamefont {Daube}, \citenamefont {Davis},\ and\ \citenamefont
  {Grainger}}]{Lin2003}%
  \BibitemOpen
  \bibfield  {author} {\bibinfo {author} {\bibfnamefont {J.~C.}\ \bibnamefont
  {Lin}}, \bibinfo {author} {\bibfnamefont {C.}~\bibnamefont {Gerbig}},
  \bibinfo {author} {\bibfnamefont {S.~C.}\ \bibnamefont {Wofsy}}, \bibinfo
  {author} {\bibfnamefont {A.~E.}\ \bibnamefont {Andrews}}, \bibinfo {author}
  {\bibfnamefont {B.~C.}\ \bibnamefont {Daube}}, \bibinfo {author}
  {\bibfnamefont {K.~J.}\ \bibnamefont {Davis}}, \ and\ \bibinfo {author}
  {\bibfnamefont {C.~A.}\ \bibnamefont {Grainger}},\ }\bibfield  {title}
  {\enquote {\bibinfo {title} {{A near-field tool for simulating the upstream
  influence of atmospheric observations: The Stochastic Time-Inverted
  Lagrangian Transport (STILT) model}},}\ }\href@noop {} {\bibfield  {journal}
  {\bibinfo  {journal} {J. Geophys. Res.}\ }\textbf {\bibinfo {volume} {108
  (D16)}},\ \bibinfo {pages} {4493} (\bibinfo {year} {2003})}\BibitemShut
  {NoStop}%
\bibitem [{\citenamefont {Szintai}, \citenamefont {Kaufmann},\ and\
  \citenamefont {Rottach}(2010)}]{Szintai2010}%
  \BibitemOpen
  \bibfield  {author} {\bibinfo {author} {\bibfnamefont {B.}~\bibnamefont
  {Szintai}}, \bibinfo {author} {\bibfnamefont {P.}~\bibnamefont {Kaufmann}}, \
  and\ \bibinfo {author} {\bibfnamefont {M.~W.}\ \bibnamefont {Rottach}},\
  }\bibfield  {title} {\enquote {\bibinfo {title} {{Simulation of pollutant
  transport in complex terrain with a numerical weather prediction–particle
  dispersion model combination}},}\ }\href@noop {} {\bibfield  {journal}
  {\bibinfo  {journal} {Boundary-Layer Meteorol.}\ }\textbf {\bibinfo {volume}
  {137}},\ \bibinfo {pages} {{373--396}} (\bibinfo {year} {2010})}\BibitemShut
  {NoStop}%
\bibitem [{\citenamefont {Sun}\ \emph {et~al.}(2018)\citenamefont {Sun},
  \citenamefont {V{\^{u}}}, \citenamefont {Ellerbroek},\ and\ \citenamefont
  {Hoekstra}}]{Sun2018}%
  \BibitemOpen
  \bibfield  {author} {\bibinfo {author} {\bibfnamefont {J.}~\bibnamefont
  {Sun}}, \bibinfo {author} {\bibfnamefont {H.}~\bibnamefont {V{\^{u}}}},
  \bibinfo {author} {\bibfnamefont {J.}~\bibnamefont {Ellerbroek}}, \ and\
  \bibinfo {author} {\bibfnamefont {J.~M.}\ \bibnamefont {Hoekstra}},\
  }\bibfield  {title} {\enquote {\bibinfo {title} {{Weather field
  reconstruction using aircraft surveillance data and a novel meteo-particle
  model}},}\ }\href@noop {} {\bibfield  {journal} {\bibinfo  {journal} {PLOS
  One}\ }\textbf {\bibinfo {volume} {13}},\ \bibinfo {pages} {{e0205029}}
  (\bibinfo {year} {2018})}\BibitemShut {NoStop}%
\bibitem [{\citenamefont {Pope}(2002)}]{Pope2002}%
  \BibitemOpen
  \bibfield  {author} {\bibinfo {author} {\bibfnamefont {S.~B.}\ \bibnamefont
  {Pope}},\ }\bibfield  {title} {\enquote {\bibinfo {title} {{Stochastic
  Lagrangian models of velocity in homogeneous turbulent shear flow}},}\
  }\href@noop {} {\bibfield  {journal} {\bibinfo  {journal} {Phys. Fluids}\
  }\textbf {\bibinfo {volume} {14}},\ \bibinfo {pages} {1696--1702} (\bibinfo
  {year} {2002})}\BibitemShut {NoStop}%
\bibitem [{\citenamefont {Li}\ \emph {et~al.}(2024)\citenamefont {Li},
  \citenamefont {Biferale}, \citenamefont {Bonaccorso}, \citenamefont
  {Scarpolini},\ and\ \citenamefont {Buzzicotti}}]{Li2024}%
  \BibitemOpen
  \bibfield  {author} {\bibinfo {author} {\bibfnamefont {T.}~\bibnamefont
  {Li}}, \bibinfo {author} {\bibfnamefont {L.}~\bibnamefont {Biferale}},
  \bibinfo {author} {\bibfnamefont {F.}~\bibnamefont {Bonaccorso}}, \bibinfo
  {author} {\bibfnamefont {M.~A.}\ \bibnamefont {Scarpolini}}, \ and\ \bibinfo
  {author} {\bibfnamefont {M.}~\bibnamefont {Buzzicotti}},\ }\bibfield  {title}
  {\enquote {\bibinfo {title} {Synthetic lagrangian turbulence by generative
  diffusion models},}\ }\href {\doibase 10.1038/s42256-024-00810-0} {\bibfield
  {journal} {\bibinfo  {journal} {Nat. Mach. Intell.}\ }\textbf {\bibinfo
  {volume} {6}},\ \bibinfo {pages} {393--403} (\bibinfo {year}
  {2024})}\BibitemShut {NoStop}%
\bibitem [{\citenamefont {Dallaire-Demers}\ and\ \citenamefont
  {Killoran}(2018)}]{Dallaire2018}%
  \BibitemOpen
  \bibfield  {author} {\bibinfo {author} {\bibfnamefont {P.-L.}\ \bibnamefont
  {Dallaire-Demers}}\ and\ \bibinfo {author} {\bibfnamefont {N.}~\bibnamefont
  {Killoran}},\ }\bibfield  {title} {\enquote {\bibinfo {title} {Quantum
  generative adversarial networks},}\ }\href {\doibase
  10.1103/PhysRevA.98.012324} {\bibfield  {journal} {\bibinfo  {journal} {Phys.
  Rev. A}\ }\textbf {\bibinfo {volume} {98}},\ \bibinfo {pages} {012324}
  (\bibinfo {year} {2018})}\BibitemShut {NoStop}%
\bibitem [{\citenamefont {Gao}\ \emph {et~al.}(2022)\citenamefont {Gao},
  \citenamefont {Anschuetz}, \citenamefont {Wang}, \citenamefont {Cirac},\ and\
  \citenamefont {Lukin}}]{Ghao2022}%
  \BibitemOpen
  \bibfield  {author} {\bibinfo {author} {\bibfnamefont {X.}~\bibnamefont
  {Gao}}, \bibinfo {author} {\bibfnamefont {E.~R.}\ \bibnamefont {Anschuetz}},
  \bibinfo {author} {\bibfnamefont {S.-T.}\ \bibnamefont {Wang}}, \bibinfo
  {author} {\bibfnamefont {J.~I.}\ \bibnamefont {Cirac}}, \ and\ \bibinfo
  {author} {\bibfnamefont {M.~D.}\ \bibnamefont {Lukin}},\ }\bibfield  {title}
  {\enquote {\bibinfo {title} {Enhancing generative models via quantum
  correlations},}\ }\href {\doibase 10.1103/PhysRevX.12.021037} {\bibfield
  {journal} {\bibinfo  {journal} {Phys. Rev. X}\ }\textbf {\bibinfo {volume}
  {12}},\ \bibinfo {pages} {021037} (\bibinfo {year} {2022})}\BibitemShut
  {NoStop}%
\bibitem [{\citenamefont {Rudolph}\ \emph {et~al.}(2022)\citenamefont
  {Rudolph}, \citenamefont {Toussaint}, \citenamefont {Katabarwa},
  \citenamefont {Johri}, \citenamefont {Peropadre},\ and\ \citenamefont
  {Perdomo-Ortiz}}]{Rudolph2022}%
  \BibitemOpen
  \bibfield  {author} {\bibinfo {author} {\bibfnamefont {M.~S.}\ \bibnamefont
  {Rudolph}}, \bibinfo {author} {\bibfnamefont {N.~B.}\ \bibnamefont
  {Toussaint}}, \bibinfo {author} {\bibfnamefont {A.}~\bibnamefont
  {Katabarwa}}, \bibinfo {author} {\bibfnamefont {S.}~\bibnamefont {Johri}},
  \bibinfo {author} {\bibfnamefont {B.}~\bibnamefont {Peropadre}}, \ and\
  \bibinfo {author} {\bibfnamefont {A.}~\bibnamefont {Perdomo-Ortiz}},\
  }\bibfield  {title} {\enquote {\bibinfo {title} {Generation of
  high-resolution handwritten digits with an ion-trap quantum computer},}\
  }\href {\doibase 10.1103/PhysRevX.12.031010} {\bibfield  {journal} {\bibinfo
  {journal} {Phys. Rev. X}\ }\textbf {\bibinfo {volume} {12}},\ \bibinfo
  {pages} {031010} (\bibinfo {year} {2022})}\BibitemShut {NoStop}%
\bibitem [{\citenamefont {Kyriienko}, \citenamefont {Paine},\ and\
  \citenamefont {Elfving}(2024)}]{Kyriienko2024}%
  \BibitemOpen
  \bibfield  {author} {\bibinfo {author} {\bibfnamefont {O.}~\bibnamefont
  {Kyriienko}}, \bibinfo {author} {\bibfnamefont {A.~E.}\ \bibnamefont
  {Paine}}, \ and\ \bibinfo {author} {\bibfnamefont {V.~E.}\ \bibnamefont
  {Elfving}},\ }\bibfield  {title} {\enquote {\bibinfo {title} {Protocols for
  trainable and differentiable quantum generative modeling},}\ }\href {\doibase
  10.1103/PhysRevResearch.6.033291} {\bibfield  {journal} {\bibinfo  {journal}
  {Phys. Rev. Research}\ }\textbf {\bibinfo {volume} {6}},\ \bibinfo {pages}
  {033291} (\bibinfo {year} {2024})}\BibitemShut {NoStop}%
\bibitem [{\citenamefont {Huang}\ \emph {et~al.}(2021)\citenamefont {Huang},
  \citenamefont {Du}, \citenamefont {Gong}, \citenamefont {Zhao}, \citenamefont
  {Wu}, \citenamefont {Wang}, \citenamefont {Li}, \citenamefont {Liang},
  \citenamefont {Lin}, \citenamefont {Xu}, \citenamefont {Yang}, \citenamefont
  {Liu}, \citenamefont {Hsieh}, \citenamefont {Deng}, \citenamefont {Rong},
  \citenamefont {Peng}, \citenamefont {Lu}, \citenamefont {Chen}, \citenamefont
  {Tao}, \citenamefont {Zhu},\ and\ \citenamefont {Pan}}]{Huang2021}%
  \BibitemOpen
  \bibfield  {author} {\bibinfo {author} {\bibfnamefont {H.-L.}\ \bibnamefont
  {Huang}}, \bibinfo {author} {\bibfnamefont {Y.}~\bibnamefont {Du}}, \bibinfo
  {author} {\bibfnamefont {M.}~\bibnamefont {Gong}}, \bibinfo {author}
  {\bibfnamefont {Y.}~\bibnamefont {Zhao}}, \bibinfo {author} {\bibfnamefont
  {Y.}~\bibnamefont {Wu}}, \bibinfo {author} {\bibfnamefont {C.}~\bibnamefont
  {Wang}}, \bibinfo {author} {\bibfnamefont {S.}~\bibnamefont {Li}}, \bibinfo
  {author} {\bibfnamefont {F.}~\bibnamefont {Liang}}, \bibinfo {author}
  {\bibfnamefont {J.}~\bibnamefont {Lin}}, \bibinfo {author} {\bibfnamefont
  {Y.}~\bibnamefont {Xu}}, \bibinfo {author} {\bibfnamefont {R.}~\bibnamefont
  {Yang}}, \bibinfo {author} {\bibfnamefont {T.}~\bibnamefont {Liu}}, \bibinfo
  {author} {\bibfnamefont {M.-H.}\ \bibnamefont {Hsieh}}, \bibinfo {author}
  {\bibfnamefont {H.}~\bibnamefont {Deng}}, \bibinfo {author} {\bibfnamefont
  {H.}~\bibnamefont {Rong}}, \bibinfo {author} {\bibfnamefont {C.-Z.}\
  \bibnamefont {Peng}}, \bibinfo {author} {\bibfnamefont {C.-Y.}\ \bibnamefont
  {Lu}}, \bibinfo {author} {\bibfnamefont {Y.-A.}\ \bibnamefont {Chen}},
  \bibinfo {author} {\bibfnamefont {D.}~\bibnamefont {Tao}}, \bibinfo {author}
  {\bibfnamefont {X.}~\bibnamefont {Zhu}}, \ and\ \bibinfo {author}
  {\bibfnamefont {J.-W.}\ \bibnamefont {Pan}},\ }\bibfield  {title} {\enquote
  {\bibinfo {title} {Experimental quantum generative adversarial networks for
  image generation},}\ }\href {\doibase 10.1103/PhysRevApplied.16.024051}
  {\bibfield  {journal} {\bibinfo  {journal} {Phys. Rev. Applied}\ }\textbf
  {\bibinfo {volume} {16}},\ \bibinfo {pages} {024051} (\bibinfo {year}
  {2021})}\BibitemShut {NoStop}%
\bibitem [{\citenamefont {Zoufal}, \citenamefont {Lucchi},\ and\ \citenamefont
  {Woerner}(2019)}]{Zoufal2019}%
  \BibitemOpen
  \bibfield  {author} {\bibinfo {author} {\bibfnamefont {C.}~\bibnamefont
  {Zoufal}}, \bibinfo {author} {\bibfnamefont {A.}~\bibnamefont {Lucchi}}, \
  and\ \bibinfo {author} {\bibfnamefont {S.}~\bibnamefont {Woerner}},\
  }\bibfield  {title} {\enquote {\bibinfo {title} {Quantum generative
  adversarial networks for learning and loading random distributions},}\ }\href
  {\doibase 10.1038/s41534-019-0223-2} {\bibfield  {journal} {\bibinfo
  {journal} {npj Quantum Inf.}\ }\textbf {\bibinfo {volume} {5}},\ \bibinfo
  {pages} {103} (\bibinfo {year} {2019})}\BibitemShut {NoStop}%
\bibitem [{\citenamefont {Romero}\ and\ \citenamefont
  {Aspuru-Guzik}(2019)}]{Romero2019}%
  \BibitemOpen
  \bibfield  {author} {\bibinfo {author} {\bibfnamefont {J.}~\bibnamefont
  {Romero}}\ and\ \bibinfo {author} {\bibfnamefont {A.}~\bibnamefont
  {Aspuru-Guzik}},\ }\bibfield  {title} {\enquote {\bibinfo {title}
  {{Variational quantum generators: Generative adversarial quantum machine
  learning for continuous distributions}},}\ }\href@noop {} {\bibfield
  {journal} {\bibinfo  {journal} {arXiv:1901.00848}\ } (\bibinfo {year}
  {2019})}\BibitemShut {NoStop}%
\bibitem [{iqm()}]{iqm2024}%
  \BibitemOpen
  \href@noop {} {}\bibinfo {note} {{IQM Resonance (2024)}, Online:
  https://www.meetiqm.com/products/iqm-resonance}\BibitemShut {NoStop}%
\bibitem [{\citenamefont {Javadi-Abhari}\ \emph {et~al.}(2024)\citenamefont
  {Javadi-Abhari}, \citenamefont {Treinish}, \citenamefont {Krsulich},
  \citenamefont {Wood}, \citenamefont {Lishman}, \citenamefont {Gacon},
  \citenamefont {Martiel}, \citenamefont {Nation}, \citenamefont {Bishop},
  \citenamefont {Cross}, \citenamefont {Johnson},\ and\ \citenamefont
  {Gambetta}}]{qiskit2024}%
  \BibitemOpen
  \bibfield  {author} {\bibinfo {author} {\bibfnamefont {A.}~\bibnamefont
  {Javadi-Abhari}}, \bibinfo {author} {\bibfnamefont {M.}~\bibnamefont
  {Treinish}}, \bibinfo {author} {\bibfnamefont {K.}~\bibnamefont {Krsulich}},
  \bibinfo {author} {\bibfnamefont {C.~J.}\ \bibnamefont {Wood}}, \bibinfo
  {author} {\bibfnamefont {J.}~\bibnamefont {Lishman}}, \bibinfo {author}
  {\bibfnamefont {J.}~\bibnamefont {Gacon}}, \bibinfo {author} {\bibfnamefont
  {S.}~\bibnamefont {Martiel}}, \bibinfo {author} {\bibfnamefont {P.~D.}\
  \bibnamefont {Nation}}, \bibinfo {author} {\bibfnamefont {L.~S.}\
  \bibnamefont {Bishop}}, \bibinfo {author} {\bibfnamefont {A.~W.}\
  \bibnamefont {Cross}}, \bibinfo {author} {\bibfnamefont {B.~R.}\ \bibnamefont
  {Johnson}}, \ and\ \bibinfo {author} {\bibfnamefont {J.~M.}\ \bibnamefont
  {Gambetta}},\ }\href {\doibase 10.48550/arXiv.2405.08810} {\enquote {\bibinfo
  {title} {Quantum computing with {Q}iskit},}\ } (\bibinfo {year} {2024}),\
  \Eprint {http://arxiv.org/abs/2405.08810} {arXiv:2405.08810 [quant-ph]}
  \BibitemShut {NoStop}%
\bibitem [{\citenamefont {Rudolph}(2020)}]{Rudolph2020}%
  \BibitemOpen
  \bibfield  {author} {\bibinfo {author} {\bibfnamefont {M.~S.}\ \bibnamefont
  {Rudolph}},\ }\href@noop {} {\emph {\bibinfo {title} {Exploring and
  Benchmarking Quantum-assisted Neural Networks with Qubit Layers}}},\ Master
  Thesis, Department of Physics and Astronomy, University of Heidelberg,
  Germany\ (\bibinfo {year} {2020})\BibitemShut {NoStop}%
\bibitem [{\citenamefont {Nielsen}\ and\ \citenamefont
  {Chuang}(2011)}]{Nielsen2010}%
  \BibitemOpen
  \bibfield  {author} {\bibinfo {author} {\bibfnamefont {M.}~\bibnamefont
  {Nielsen}}\ and\ \bibinfo {author} {\bibfnamefont {I.}~\bibnamefont
  {Chuang}},\ }\href@noop {} {\emph {\bibinfo {title} {Quantum Computation and
  Quantum Information: 10th Anniversary Edition}}}\ (\bibinfo  {publisher}
  {Cambridge University Press},\ \bibinfo {year} {2011})\BibitemShut {NoStop}%
\bibitem [{\citenamefont {Holmes}\ \emph {et~al.}(2022)\citenamefont {Holmes},
  \citenamefont {Sharma}, \citenamefont {Cerezo},\ and\ \citenamefont
  {Coles}}]{Holmes2022}%
  \BibitemOpen
  \bibfield  {author} {\bibinfo {author} {\bibfnamefont {Z.}~\bibnamefont
  {Holmes}}, \bibinfo {author} {\bibfnamefont {K.}~\bibnamefont {Sharma}},
  \bibinfo {author} {\bibfnamefont {M.}~\bibnamefont {Cerezo}}, \ and\ \bibinfo
  {author} {\bibfnamefont {P.~J.}\ \bibnamefont {Coles}},\ }\bibfield  {title}
  {\enquote {\bibinfo {title} {{Connecting Ansatz Expressibility to Gradient
  Magnitudes and Barren Plateaus}},}\ }\href {\doibase
  10.1103/PRXQuantum.3.010313} {\bibfield  {journal} {\bibinfo  {journal} {PRX
  Quantum}\ }\textbf {\bibinfo {volume} {3}},\ \bibinfo {pages} {010313}
  (\bibinfo {year} {2022})}\BibitemShut {NoStop}%
\bibitem [{\citenamefont {Nakhl}, \citenamefont {Quella},\ and\ \citenamefont
  {Usman}(2024)}]{Nakhl2024}%
  \BibitemOpen
  \bibfield  {author} {\bibinfo {author} {\bibfnamefont {A.~C.}\ \bibnamefont
  {Nakhl}}, \bibinfo {author} {\bibfnamefont {A.~C.}\ \bibnamefont {Quella}}, \
  and\ \bibinfo {author} {\bibfnamefont {M.}~\bibnamefont {Usman}},\ }\bibfield
   {title} {\enquote {\bibinfo {title} {Calibrating the role of entanglement in
  variational quantum circuits},}\ }\href {\doibase
  10.1103/PhysRevA.109.032413} {\bibfield  {journal} {\bibinfo  {journal}
  {Phys. Rev. A}\ }\textbf {\bibinfo {volume} {109}},\ \bibinfo {pages}
  {032413} (\bibinfo {year} {2024})}\BibitemShut {NoStop}%
\bibitem [{\citenamefont {Nelder}\ and\ \citenamefont
  {Mead}(1965)}]{Nelder1965}%
  \BibitemOpen
  \bibfield  {author} {\bibinfo {author} {\bibfnamefont {J.}~\bibnamefont
  {Nelder}}\ and\ \bibinfo {author} {\bibfnamefont {R.}~\bibnamefont {Mead}},\
  }\bibfield  {title} {\enquote {\bibinfo {title} {A simplex method for
  function minimization},}\ }\href {\doibase 10.1093/comjnl/7.4.308} {\bibfield
   {journal} {\bibinfo  {journal} {The Computer Journal}\ }\textbf {\bibinfo
  {volume} {7}},\ \bibinfo {pages} {308--313} (\bibinfo {year}
  {1965})}\BibitemShut {NoStop}%
\bibitem [{\citenamefont {Benedetti}\ \emph {et~al.}(2020)\citenamefont
  {Benedetti}, \citenamefont {Garcia-Pintos}, \citenamefont {Perdomo},
  \citenamefont {Leyton-Ortega}, \citenamefont {Nam},\ and\ \citenamefont
  {Perdomo-Ortiz}}]{Benedetti2019}%
  \BibitemOpen
  \bibfield  {author} {\bibinfo {author} {\bibfnamefont {M.}~\bibnamefont
  {Benedetti}}, \bibinfo {author} {\bibfnamefont {D.}~\bibnamefont
  {Garcia-Pintos}}, \bibinfo {author} {\bibfnamefont {O.}~\bibnamefont
  {Perdomo}}, \bibinfo {author} {\bibfnamefont {V.}~\bibnamefont
  {Leyton-Ortega}}, \bibinfo {author} {\bibfnamefont {Y.}~\bibnamefont {Nam}},
  \ and\ \bibinfo {author} {\bibfnamefont {A.}~\bibnamefont {Perdomo-Ortiz}},\
  }\bibfield  {title} {\enquote {\bibinfo {title} {{A generative modeling
  approach for benchmarking and training shallow quantum circuits}},}\
  }\href@noop {} {\bibfield  {journal} {\bibinfo  {journal} {npj Quantum Inf.}\
  }\textbf {\bibinfo {volume} {5}},\ \bibinfo {pages} {45} (\bibinfo {year}
  {2020})}\BibitemShut {NoStop}%
\bibitem [{\citenamefont {Pope}(2000)}]{Pope2000}%
  \BibitemOpen
  \bibfield  {author} {\bibinfo {author} {\bibfnamefont {S.~B.}\ \bibnamefont
  {Pope}},\ }\href@noop {} {\emph {\bibinfo {title} {Turbulent Flows}}}\
  (\bibinfo  {publisher} {Cambridge University Press},\ \bibinfo {address}
  {Cambridge, UK},\ \bibinfo {year} {2000})\BibitemShut {NoStop}%
\bibitem [{\citenamefont {Rogallo}(1986)}]{Rogallo1981}%
  \BibitemOpen
  \bibfield  {author} {\bibinfo {author} {\bibfnamefont {R.~S.}\ \bibnamefont
  {Rogallo}},\ }\href@noop {} {\emph {\bibinfo {title} {Numerical experiments
  in homogeneous turbulence}}},\ NASA Technical Memorandum No. 81315\ (\bibinfo
  {year} {1986})\BibitemShut {NoStop}%
\bibitem [{\citenamefont {Rogers}, \citenamefont {Moin},\ and\ \citenamefont
  {Reynolds}(1986)}]{Rogers1986}%
  \BibitemOpen
  \bibfield  {author} {\bibinfo {author} {\bibfnamefont {M.~M.}\ \bibnamefont
  {Rogers}}, \bibinfo {author} {\bibfnamefont {P.}~\bibnamefont {Moin}}, \ and\
  \bibinfo {author} {\bibfnamefont {W.~C.}\ \bibnamefont {Reynolds}},\
  }\href@noop {} {\emph {\bibinfo {title} {The Structure and Modeling of the
  Hydrodynamic and Passive Scalar Fields in Homogeneous Turbulent Shear
  Flow}}},\ Report No. TF-25\ (\bibinfo  {publisher} {Department of Mechanical
  Engineering, Stanford University},\ \bibinfo {year} {1986})\BibitemShut
  {NoStop}%
\bibitem [{\citenamefont {Pumir}(1996)}]{Pumir1996}%
  \BibitemOpen
  \bibfield  {author} {\bibinfo {author} {\bibfnamefont {A.}~\bibnamefont
  {Pumir}},\ }\bibfield  {title} {\enquote {\bibinfo {title} {Turbulence in
  homogeneous shear flows},}\ }\href@noop {} {\bibfield  {journal} {\bibinfo
  {journal} {Phys. Fluids}\ }\textbf {\bibinfo {volume} {8}},\ \bibinfo {pages}
  {3112--3127} (\bibinfo {year} {1996})}\BibitemShut {NoStop}%
\bibitem [{\citenamefont {Schumacher}\ and\ \citenamefont
  {Eckhardt}(2000)}]{Schumacher2000}%
  \BibitemOpen
  \bibfield  {author} {\bibinfo {author} {\bibfnamefont {J.}~\bibnamefont
  {Schumacher}}\ and\ \bibinfo {author} {\bibfnamefont {B.}~\bibnamefont
  {Eckhardt}},\ }\bibfield  {title} {\enquote {\bibinfo {title} {On
  statistically stationary homogeneous shear turbulence},}\ }\href@noop {}
  {\bibfield  {journal} {\bibinfo  {journal} {Europhys. Lett.}\ }\textbf
  {\bibinfo {volume} {52}},\ \bibinfo {pages} {627--632} (\bibinfo {year}
  {2000})}\BibitemShut {NoStop}%
\bibitem [{\citenamefont {Gualtieri}\ \emph {et~al.}(2002)\citenamefont
  {Gualtieri}, \citenamefont {Casciola}, \citenamefont {Benzi}, \citenamefont
  {Amati},\ and\ \citenamefont {Piva}}]{Gualtieri2002}%
  \BibitemOpen
  \bibfield  {author} {\bibinfo {author} {\bibfnamefont {P.}~\bibnamefont
  {Gualtieri}}, \bibinfo {author} {\bibfnamefont {C.~M.}\ \bibnamefont
  {Casciola}}, \bibinfo {author} {\bibfnamefont {R.}~\bibnamefont {Benzi}},
  \bibinfo {author} {\bibfnamefont {G.}~\bibnamefont {Amati}}, \ and\ \bibinfo
  {author} {\bibfnamefont {R.}~\bibnamefont {Piva}},\ }\bibfield  {title}
  {\enquote {\bibinfo {title} {Scaling laws and intermittency in homogeneous
  shear flow},}\ }\href@noop {} {\bibfield  {journal} {\bibinfo  {journal}
  {Phys. Fluids}\ }\textbf {\bibinfo {volume} {14}},\ \bibinfo {pages}
  {583--596} (\bibinfo {year} {2002})}\BibitemShut {NoStop}%
\bibitem [{\citenamefont {Schumacher}(2004)}]{Schumacher2004}%
  \BibitemOpen
  \bibfield  {author} {\bibinfo {author} {\bibfnamefont {J.}~\bibnamefont
  {Schumacher}},\ }\bibfield  {title} {\enquote {\bibinfo {title} {{Relation
  between shear parameter and Reynolds number in statistically stationary
  turbulent shear flows}},}\ }\href@noop {} {\bibfield  {journal} {\bibinfo
  {journal} {Phys. Fluids}\ }\textbf {\bibinfo {volume} {16}},\ \bibinfo
  {pages} {3094--3102} (\bibinfo {year} {2004})}\BibitemShut {NoStop}%
\bibitem [{\citenamefont {Yakhot}(2003)}]{Yakhot2003}%
  \BibitemOpen
  \bibfield  {author} {\bibinfo {author} {\bibfnamefont {V.}~\bibnamefont
  {Yakhot}},\ }\bibfield  {title} {\enquote {\bibinfo {title} {{A simple model
  for self-sustained oscillations in homogeneous shear flow}},}\ }\href@noop {}
  {\bibfield  {journal} {\bibinfo  {journal} {Phys. Fluids}\ }\textbf {\bibinfo
  {volume} {15}},\ \bibinfo {pages} {{L17--L20}} (\bibinfo {year}
  {2003})}\BibitemShut {NoStop}%
\bibitem [{\citenamefont {Higham}\ and\ \citenamefont
  {Kloeden}(2021)}]{Higham2021}%
  \BibitemOpen
  \bibfield  {author} {\bibinfo {author} {\bibfnamefont {D.~J.}\ \bibnamefont
  {Higham}}\ and\ \bibinfo {author} {\bibfnamefont {P.~E.}\ \bibnamefont
  {Kloeden}},\ }\href@noop {} {\emph {\bibinfo {title} {An Introduction to the
  Numerical Simulation of Stochastic Differential Equations}}}\ (\bibinfo
  {publisher} {SIAM, Philadelphia},\ \bibinfo {year} {2021})\BibitemShut
  {NoStop}%
\bibitem [{\citenamefont {Köcher}\ \emph {et~al.}(2025)\citenamefont
  {Köcher}, \citenamefont {Rose}, \citenamefont {Bharadwaj}, \citenamefont
  {Schumacher},\ and\ \citenamefont {Schumacher}}]{Koecher2025}%
  \BibitemOpen
  \bibfield  {author} {\bibinfo {author} {\bibfnamefont {N.}~\bibnamefont
  {Köcher}}, \bibinfo {author} {\bibfnamefont {H.}~\bibnamefont {Rose}},
  \bibinfo {author} {\bibfnamefont {S.~S.}\ \bibnamefont {Bharadwaj}}, \bibinfo
  {author} {\bibfnamefont {J.}~\bibnamefont {Schumacher}}, \ and\ \bibinfo
  {author} {\bibfnamefont {S.}~\bibnamefont {Schumacher}},\ }\bibfield  {title}
  {\enquote {\bibinfo {title} {{Numerical solution of nonlinear Schrödinger
  equation by a hybrid pseudospectral-variational quantum algorithm}},}\
  }\href@noop {} {\bibfield  {journal} {\bibinfo  {journal} {Sci. Rep.}\
  }\textbf {\bibinfo {volume} {15}},\ \bibinfo {pages} {{in press}} (\bibinfo
  {year} {2025})}\BibitemShut {NoStop}%
\bibitem [{\citenamefont {Marxer}\ \emph {et~al.}(2023)\citenamefont {Marxer},
  \citenamefont {Veps\"al\"ainen}, \citenamefont {Jolin}, \citenamefont
  {Tuorila}, \citenamefont {Landra}, \citenamefont {Ockeloen-Korppi},
  \citenamefont {Liu}, \citenamefont {Ahonen}, \citenamefont {Auer},
  \citenamefont {Belzane}, \citenamefont {Bergholm}, \citenamefont {Chan},
  \citenamefont {Chan}, \citenamefont {Hiltunen}, \citenamefont {Hotari},
  \citenamefont {Hyypp\"a}, \citenamefont {Ikonen}, \citenamefont {Janzso},
  \citenamefont {Koistinen}, \citenamefont {Kotilahti}, \citenamefont {Li},
  \citenamefont {Luus}, \citenamefont {Papic}, \citenamefont {Partanen},
  \citenamefont {R\"abin\"a}, \citenamefont {Rosti}, \citenamefont {Savytskyi},
  \citenamefont {Sepp\"al\"a}, \citenamefont {Sevriuk}, \citenamefont {Takala},
  \citenamefont {Tarasinski}, \citenamefont {Thapa}, \citenamefont {Tosto},
  \citenamefont {Vorobeva}, \citenamefont {Yu}, \citenamefont {Tan},
  \citenamefont {Hassel}, \citenamefont {M\"ott\"onen},\ and\ \citenamefont
  {Heinsoo}}]{Marxer2023}%
  \BibitemOpen
  \bibfield  {author} {\bibinfo {author} {\bibfnamefont {F.}~\bibnamefont
  {Marxer}}, \bibinfo {author} {\bibfnamefont {A.}~\bibnamefont
  {Veps\"al\"ainen}}, \bibinfo {author} {\bibfnamefont {S.~W.}\ \bibnamefont
  {Jolin}}, \bibinfo {author} {\bibfnamefont {J.}~\bibnamefont {Tuorila}},
  \bibinfo {author} {\bibfnamefont {A.}~\bibnamefont {Landra}}, \bibinfo
  {author} {\bibfnamefont {C.}~\bibnamefont {Ockeloen-Korppi}}, \bibinfo
  {author} {\bibfnamefont {W.}~\bibnamefont {Liu}}, \bibinfo {author}
  {\bibfnamefont {O.}~\bibnamefont {Ahonen}}, \bibinfo {author} {\bibfnamefont
  {A.}~\bibnamefont {Auer}}, \bibinfo {author} {\bibfnamefont {L.}~\bibnamefont
  {Belzane}}, \bibinfo {author} {\bibfnamefont {V.}~\bibnamefont {Bergholm}},
  \bibinfo {author} {\bibfnamefont {C.~F.}\ \bibnamefont {Chan}}, \bibinfo
  {author} {\bibfnamefont {K.~W.}\ \bibnamefont {Chan}}, \bibinfo {author}
  {\bibfnamefont {T.}~\bibnamefont {Hiltunen}}, \bibinfo {author}
  {\bibfnamefont {J.}~\bibnamefont {Hotari}}, \bibinfo {author} {\bibfnamefont
  {E.}~\bibnamefont {Hyypp\"a}}, \bibinfo {author} {\bibfnamefont
  {J.}~\bibnamefont {Ikonen}}, \bibinfo {author} {\bibfnamefont
  {D.}~\bibnamefont {Janzso}}, \bibinfo {author} {\bibfnamefont
  {M.}~\bibnamefont {Koistinen}}, \bibinfo {author} {\bibfnamefont
  {J.}~\bibnamefont {Kotilahti}}, \bibinfo {author} {\bibfnamefont
  {T.}~\bibnamefont {Li}}, \bibinfo {author} {\bibfnamefont {J.}~\bibnamefont
  {Luus}}, \bibinfo {author} {\bibfnamefont {M.}~\bibnamefont {Papic}},
  \bibinfo {author} {\bibfnamefont {M.}~\bibnamefont {Partanen}}, \bibinfo
  {author} {\bibfnamefont {J.}~\bibnamefont {R\"abin\"a}}, \bibinfo {author}
  {\bibfnamefont {J.}~\bibnamefont {Rosti}}, \bibinfo {author} {\bibfnamefont
  {M.}~\bibnamefont {Savytskyi}}, \bibinfo {author} {\bibfnamefont
  {M.}~\bibnamefont {Sepp\"al\"a}}, \bibinfo {author} {\bibfnamefont
  {V.}~\bibnamefont {Sevriuk}}, \bibinfo {author} {\bibfnamefont
  {E.}~\bibnamefont {Takala}}, \bibinfo {author} {\bibfnamefont
  {B.}~\bibnamefont {Tarasinski}}, \bibinfo {author} {\bibfnamefont {M.~J.}\
  \bibnamefont {Thapa}}, \bibinfo {author} {\bibfnamefont {F.}~\bibnamefont
  {Tosto}}, \bibinfo {author} {\bibfnamefont {N.}~\bibnamefont {Vorobeva}},
  \bibinfo {author} {\bibfnamefont {L.}~\bibnamefont {Yu}}, \bibinfo {author}
  {\bibfnamefont {K.~Y.}\ \bibnamefont {Tan}}, \bibinfo {author} {\bibfnamefont
  {J.}~\bibnamefont {Hassel}}, \bibinfo {author} {\bibfnamefont
  {M.}~\bibnamefont {M\"ott\"onen}}, \ and\ \bibinfo {author} {\bibfnamefont
  {J.}~\bibnamefont {Heinsoo}},\ }\bibfield  {title} {\enquote {\bibinfo
  {title} {{Long-distance transmon coupler with CZ-gate fidelity above
  $99.8\%$}},}\ }\href {\doibase 10.1103/PRXQuantum.4.010314} {\bibfield
  {journal} {\bibinfo  {journal} {PRX Quantum}\ }\textbf {\bibinfo {volume}
  {4}},\ \bibinfo {pages} {010314} (\bibinfo {year} {2023})}\BibitemShut
  {NoStop}%
\bibitem [{Qis(2025)}]{Qiskit}%
  \BibitemOpen
  \bibfield  {title} {\enquote {\bibinfo {title} {Qiskit version 1.1.2},}\
  }\href@noop {} {\  (\bibinfo {year} {2025})}\BibitemShut {NoStop}%
\bibitem [{\citenamefont {Sawford}\ and\ \citenamefont
  {Yeung}(2000)}]{Sawford2000}%
  \BibitemOpen
  \bibfield  {author} {\bibinfo {author} {\bibfnamefont {B.~L.}\ \bibnamefont
  {Sawford}}\ and\ \bibinfo {author} {\bibfnamefont {P.~K.}\ \bibnamefont
  {Yeung}},\ }\bibfield  {title} {\enquote {\bibinfo {title} {{Eulerian
  acceleration statistics as a discriminator between Lagrangian stochastic
  models in uniform shear flow}},}\ }\href@noop {} {\bibfield  {journal}
  {\bibinfo  {journal} {Phys. Fluids}\ }\textbf {\bibinfo {volume} {12}},\
  \bibinfo {pages} {2033--2045} (\bibinfo {year} {2000})}\BibitemShut {NoStop}%
\bibitem [{\citenamefont {Sawford}\ and\ \citenamefont
  {Yeung}(2001)}]{Sawford2001}%
  \BibitemOpen
  \bibfield  {author} {\bibinfo {author} {\bibfnamefont {B.~L.}\ \bibnamefont
  {Sawford}}\ and\ \bibinfo {author} {\bibfnamefont {P.~K.}\ \bibnamefont
  {Yeung}},\ }\bibfield  {title} {\enquote {\bibinfo {title} {{Lagrangian
  statistics in uniform shear flow: Direct numerical simulation and Lagrangian
  stochastic models}},}\ }\href@noop {} {\bibfield  {journal} {\bibinfo
  {journal} {Phys. Fluids}\ }\textbf {\bibinfo {volume} {13}},\ \bibinfo
  {pages} {2627--2634} (\bibinfo {year} {2001})}\BibitemShut {NoStop}%
\bibitem [{\citenamefont {Shen}\ and\ \citenamefont {Yeung}(1997)}]{Shen1997}%
  \BibitemOpen
  \bibfield  {author} {\bibinfo {author} {\bibfnamefont {P.}~\bibnamefont
  {Shen}}\ and\ \bibinfo {author} {\bibfnamefont {P.~K.}\ \bibnamefont
  {Yeung}},\ }\bibfield  {title} {\enquote {\bibinfo {title} {{Fluid particle
  dispersion in homogeneous turbulent shear flow}},}\ }\href@noop {} {\bibfield
   {journal} {\bibinfo  {journal} {Phys. Fluids}\ }\textbf {\bibinfo {volume}
  {9}},\ \bibinfo {pages} {3472--3483} (\bibinfo {year} {1997})}\BibitemShut
  {NoStop}%
\bibitem [{\citenamefont {Boffetta}\ and\ \citenamefont
  {Sokolov}(2002)}]{Boffetta2002}%
  \BibitemOpen
  \bibfield  {author} {\bibinfo {author} {\bibfnamefont {G.}~\bibnamefont
  {Boffetta}}\ and\ \bibinfo {author} {\bibfnamefont {I.~M.}\ \bibnamefont
  {Sokolov}},\ }\bibfield  {title} {\enquote {\bibinfo {title} {{Statistics of
  two-particle dispersion in two-dimensional turbulence}},}\ }\href@noop {}
  {\bibfield  {journal} {\bibinfo  {journal} {Phys. Fluids}\ }\textbf {\bibinfo
  {volume} {14}},\ \bibinfo {pages} {3224--3232} (\bibinfo {year}
  {2002})}\BibitemShut {NoStop}%
\bibitem [{\citenamefont {Biferale}\ \emph {et~al.}(2005)\citenamefont
  {Biferale}, \citenamefont {Boffetta}, \citenamefont {Celani}, \citenamefont
  {Devenisch}, \citenamefont {Lanotte},\ and\ \citenamefont
  {Toschi}}]{Biferale2005}%
  \BibitemOpen
  \bibfield  {author} {\bibinfo {author} {\bibfnamefont {L.}~\bibnamefont
  {Biferale}}, \bibinfo {author} {\bibfnamefont {G.}~\bibnamefont {Boffetta}},
  \bibinfo {author} {\bibfnamefont {A.}~\bibnamefont {Celani}}, \bibinfo
  {author} {\bibfnamefont {B.~J.}\ \bibnamefont {Devenisch}}, \bibinfo {author}
  {\bibfnamefont {A.}~\bibnamefont {Lanotte}}, \ and\ \bibinfo {author}
  {\bibfnamefont {F.}~\bibnamefont {Toschi}},\ }\bibfield  {title} {\enquote
  {\bibinfo {title} {{Lagrangian statistics of particle pairs in homogeneous
  isotropic turbulence}},}\ }\href@noop {} {\bibfield  {journal} {\bibinfo
  {journal} {Phys. Fluids}\ }\textbf {\bibinfo {volume} {17}},\ \bibinfo
  {pages} {115101} (\bibinfo {year} {2005})}\BibitemShut {NoStop}%
\bibitem [{\citenamefont {Schlatter}\ \emph {et~al.}(2010)\citenamefont
  {Schlatter}, \citenamefont {Li}, \citenamefont {Brethouwer}, \citenamefont
  {Johansson},\ and\ \citenamefont {Henningson}}]{Schlatter2010}%
  \BibitemOpen
  \bibfield  {author} {\bibinfo {author} {\bibfnamefont {P.}~\bibnamefont
  {Schlatter}}, \bibinfo {author} {\bibfnamefont {Q.}~\bibnamefont {Li}},
  \bibinfo {author} {\bibfnamefont {G.}~\bibnamefont {Brethouwer}}, \bibinfo
  {author} {\bibfnamefont {A.~V.}\ \bibnamefont {Johansson}}, \ and\ \bibinfo
  {author} {\bibfnamefont {D.~S.}\ \bibnamefont {Henningson}},\ }\bibfield
  {title} {\enquote {\bibinfo {title} {{Simulations of spatially evolving
  turbulent boundary layers up to $Re_{\theta}=4300$}},}\ }\href@noop {}
  {\bibfield  {journal} {\bibinfo  {journal} {Int. J. Heat Fluid Flow}\
  }\textbf {\bibinfo {volume} {31}},\ \bibinfo {pages} {251--261} (\bibinfo
  {year} {2010})}\BibitemShut {NoStop}%
\bibitem [{\citenamefont {Layden}\ \emph {et~al.}(2023)\citenamefont {Layden},
  \citenamefont {Mazzola}, \citenamefont {Mishmash}, \citenamefont {Motta},
  \citenamefont {Wocjan}, \citenamefont {Kim},\ and\ \citenamefont
  {Sheldon}}]{Layden2023}%
  \BibitemOpen
  \bibfield  {author} {\bibinfo {author} {\bibfnamefont {D.}~\bibnamefont
  {Layden}}, \bibinfo {author} {\bibfnamefont {G.}~\bibnamefont {Mazzola}},
  \bibinfo {author} {\bibfnamefont {R.~V.}\ \bibnamefont {Mishmash}}, \bibinfo
  {author} {\bibfnamefont {M.}~\bibnamefont {Motta}}, \bibinfo {author}
  {\bibfnamefont {P.}~\bibnamefont {Wocjan}}, \bibinfo {author} {\bibfnamefont
  {J.-S.}\ \bibnamefont {Kim}}, \ and\ \bibinfo {author} {\bibfnamefont
  {S.}~\bibnamefont {Sheldon}},\ }\bibfield  {title} {\enquote {\bibinfo
  {title} {{Quantum-enhanced Markov-chain Monte Carlo}},}\ }\href@noop {}
  {\bibfield  {journal} {\bibinfo  {journal} {Nature}\ }\textbf {\bibinfo
  {volume} {619}},\ \bibinfo {pages} {282--287} (\bibinfo {year}
  {2023})}\BibitemShut {NoStop}%
\bibitem [{\citenamefont {Taylor}(1921)}]{Taylor1921}%
  \BibitemOpen
  \bibfield  {author} {\bibinfo {author} {\bibfnamefont {G.~I.}\ \bibnamefont
  {Taylor}},\ }\bibfield  {title} {\enquote {\bibinfo {title} {{Diffusion by
  continuous movements}},}\ }\href@noop {} {\bibfield  {journal} {\bibinfo
  {journal} {Proc. Lond. Math. Soc., Series 2}\ }\textbf {\bibinfo {volume}
  {20}},\ \bibinfo {pages} {196--212} (\bibinfo {year} {1921})}\BibitemShut
  {NoStop}%
\bibitem [{\citenamefont {Haworth}\ and\ \citenamefont
  {Pope}(1986)}]{Haworth1986}%
  \BibitemOpen
  \bibfield  {author} {\bibinfo {author} {\bibfnamefont {D.~C.}\ \bibnamefont
  {Haworth}}\ and\ \bibinfo {author} {\bibfnamefont {S.~B.}\ \bibnamefont
  {Pope}},\ }\bibfield  {title} {\enquote {\bibinfo {title} {{A generalized
  Langevin model for turbulent flows}},}\ }\href@noop {} {\bibfield  {journal}
  {\bibinfo  {journal} {Phys. Fluids}\ }\textbf {\bibinfo {volume} {29}},\
  \bibinfo {pages} {387--405} (\bibinfo {year} {1986})}\BibitemShut {NoStop}%
\bibitem [{\citenamefont {Thomson}(1987)}]{Thomson1987}%
  \BibitemOpen
  \bibfield  {author} {\bibinfo {author} {\bibfnamefont {D.~J.}\ \bibnamefont
  {Thomson}},\ }\bibfield  {title} {\enquote {\bibinfo {title} {{Criteria for
  the selection of stochastic models of particle trajectories in turbulent
  flows}},}\ }\href@noop {} {\bibfield  {journal} {\bibinfo  {journal} {J.
  Fluid Mech.}\ }\textbf {\bibinfo {volume} {180}},\ \bibinfo {pages}
  {529--556} (\bibinfo {year} {1987})}\BibitemShut {NoStop}%
\bibitem [{\citenamefont {Wilson}, \citenamefont {Flesch},\ and\ \citenamefont
  {Swaters}(1993)}]{Wilson1993}%
  \BibitemOpen
  \bibfield  {author} {\bibinfo {author} {\bibfnamefont {J.~D.}\ \bibnamefont
  {Wilson}}, \bibinfo {author} {\bibfnamefont {T.~K.}\ \bibnamefont {Flesch}},
  \ and\ \bibinfo {author} {\bibfnamefont {G.~E.}\ \bibnamefont {Swaters}},\
  }\bibfield  {title} {\enquote {\bibinfo {title} {{Dispersion in sheared
  Gaussian homogeneous turbulence}},}\ }\href@noop {} {\bibfield  {journal}
  {\bibinfo  {journal} {Boundary-Layer Meteorol.}\ }\textbf {\bibinfo {volume}
  {62}},\ \bibinfo {pages} {281--290} (\bibinfo {year} {1993})}\BibitemShut
  {NoStop}%
\bibitem [{\citenamefont {Bishop}(2006)}]{Bishop2006}%
  \BibitemOpen
  \bibfield  {author} {\bibinfo {author} {\bibfnamefont {C.~M.}\ \bibnamefont
  {Bishop}},\ }\href@noop {} {\emph {\bibinfo {title} {Pattern recognition and
  machine learning}}}\ (\bibinfo  {publisher} {Springer, New York},\ \bibinfo
  {year} {2006})\BibitemShut {NoStop}%
\end{thebibliography}%
\end{document}